\documentclass[12pt,reqno]{article}
\usepackage{amsmath}
\usepackage{amsthm}
\usepackage{graphicx,indentfirst}
\usepackage[psamsfonts]{amssymb}
\numberwithin{equation}{section}

\def\Lcd{h}
\def\be{\begin{equation}}
\def\ee{\end{equation}}
\def\a{\alpha}
\def\bLcd{\bar \Lcd}
\def\bw{\bar w}
\def\nn{\nonumber}
\def\QR{\mathbb{R}}
\def\Y{\Upsilon}
\def\t{\tau}
\def\cP{{\cal P}}
\def \Fig#1#2#3 {
\begin{figure}
\centering
\epsfxsize=#2cm \epsfbox{#1.eps}
\caption{#3}
\label{#1}
\end{figure}
}

\newcount\figno
\def\fig#1#2#3{
\par\begingroup\parindent=0pt\leftskip=1cm\rightskip=1cm\parindent=0pt
\baselineskip=15pt
\global\advance\figno by 1
\epsfxsize=#3
\centerline{\epsfbox{#2}}
\vskip 12pt
{\bf \small Figure \the\figno:} {\small #1}\par
\endgroup\par
}
\def\figlabel#1{\xdef#1{\the\figno
\mbox{ }}}
\def\encadremath#1{\vbox{\hrule\hbox{\vrule\kern8pt\vbox{\kern8pt
\hbox{$\displaystyle #1$}\kern8pt}
\kern8pt\vrule}\hrule}}

\def\b{{\beta}}

\def\c{{\gamma}}

\def\pl{\partial}

\def\bz{\bar{z}}

\def\ppe{\hspace*{-2.5mm}}
\def\ew{\hspace*{-1mm}}
\newcommand{\Fus}[6]{F_{{\scriptstyle #1}{\scriptstyle #2}}
  \hspace*{.3mm}\displaystyle{[} \ew \begin{array}{ll} {\scriptstyle #3 }
  \ppe & {\scriptstyle #4} \ppe \\[-2mm] {\scriptstyle #5}\ppe &
  {\scriptstyle #6}\ew \end{array}\displaystyle{]}}

\def\Vb{\mathbb{V}}
\def\bea{\begin{eqnarray*}}
\def\eea{\end{eqnarray*}}
\def\talpha{
\alpha}
\def\cN{${\mathcal N}$}

\DeclareMathAlphabet{\mathscr}{OT1}{pzc}{m}{it}
\def\L{\mathcal{L}}
\def\mcV{\mathcal V}
\def\D{\Delta}
\def\bhtwo{\hat b^{(2)}}
\def\halpha{{\hat \alpha}}

\title{Liouville's Imaginary Shadow}
\author{Volker Schomerus$^a$  and Paulina Suchanek$^{a,b}$
\\[5mm]
$^a$DESY Theory Group, DESY Hamburg  \\
$\,$Notkestrasse 85, D-22603 Hamburg, Germany
\\
\\
$^b$Institute for Theoretical Physics, University of Wroc{\l}aw, \\
$\;$pl.~M.~Borna 9, 50-204 Wroc{\l}aw, Poland
}

\date{Oct 2012}

\begin{document}
\begin{titlepage}      \maketitle      \thispagestyle{empty}

\vskip1cm
\begin{abstract}
N=1 super Liouville field theory is one of the simplest non-rational conformal
field theories. It possesses various important extensions and interesting
applications, e.g.\ to the AGT relation with 4D gauge theory or the
construction of the OSP(1$|$2) WZW model. In both setups, the N=1 Liouville
field is accompanied by an additional free fermion. Recently, Belavin
et al. suggested a bosonization of the product theory in terms of two bosonic
Liouville fields. While one of these Liouville fields is standard,
the second turns out to be imaginary (or time-like). We extend the proposal
to the R sector and perform extensive checks based on detailed comparison of
3-point functions involving several super-conformal primaries and descendants.
On the basis of such strong evidence we sketch a number of interesting
potential applications of this intriguing bozonization.
\end{abstract}

\vspace*{-16.9cm}\noindent 
{\tt {DESY 12-165}}
\bigskip\vfill
\noindent
\phantom{wwwx}{\small e-mail: }{\small\tt
paulina@ift.uni.wroc.pl, volker.schomerus@desy.de}
\end{titlepage}

\baselineskip=19pt


\setcounter{equation}{0}
\tableofcontents
\newpage

\section{Introduction}
In this work we consider an interesting bosonization of \cN=1
Liouville field theory that was proposed recently in
\cite{Belavin:2011sw}. \cN=1 Liouville field theory contains one
fermionic field $\psi$ in addition to the Liouville field
$\varphi$. These fields are coupled through the standard
interaction term. For bosonization we need to add another
free fermion $\eta$. The product theory appears naturally
in several applications of \cN=1 Liouville field theory. In
particular, it has been used in \cite{Hikida:2007sz} and
\cite{Creutzig:2010zp} to compute various structure constants
of the OSP(1$|$2) WZW model. More recently, it was considered
in the context of the AGT correspondence \cite{Alday:2009aq}
between supersymmetric 4D gauge theories and 2D conformal
field theory \cite{Belavin:2011pp, Nishioka:2011jk, Bonelli:2011jx, Bonelli:2011kv, Belavin:2011sw}.

In the bosonization, the two fermionic fields $\psi$ and
$\eta$ are replaced by a single boson $Y$. What Belavin et
al. proposed was that the two bosonic fields $\varphi$ and
$Y$ can me mapped to a new set of bosonic fields, $X$ and
$\hat X$, where $X$ is an ordinary (non-suspersymmetric)
Liouville field and $\hat X$ an imaginary cousin. The latter
may be thought of as a Liouville field which takes values
in imaginary numbers. Because of its internal structure,
we shall often refer to the fully bosonic model as {\em
double Liouville theory} and to the factor associated
with the field $\hat X$  as imaginary Liouville theory.

Imaginary Liouville theory is far from being
an established model of 2-dimensional conformal field theory.
In fact, there exist several different proposals for its
structure constants but consistency (crossing symmetry)
has never been established (see discussion in section 3). It
is remarkable that one version of imaginary Liouville theory
now appears through the bosonization of a consistent local
conformal field theory.

The relation between \cN=1 and double Liouville theory has
a suggestive ancestor in rational conformal field theory.
In that context, double Liouville theory gets replaced by a
product of two minimal models and \cN=1 Liouville theory by
its rational counterpart. We can give a highly suggestive
argument for their relation if we represent both models as
coset conformal field theories. It is well known that
ordinary minimal models arise through the cosets
$$ {\text{MM}}_k = (SU(2)_{k} \times SU(2)_1)
/ SU(2)_{k+1}
$$
where $ k = 1, 2, \dots$. This family of rational models
includes the Ising model MM$_1$ for a single fermion
$\eta$ when $k=1$. Similarly, \cN=1 supersymmetric minimal models
are obtained from the coset
$$ {\text{SMM}}_k = (SU(2)_k \times SU(2)_2)/SU(2)_{k+2}\ .
$$
If we allow ourselves to extend and reduce both numerator
and denominator by the required additional factors we
can easily see that
\begin{equation} \label{mainrat}
{\text{SMM}}_{k-1} \times {\text{MM}}_1  \sim
{\text{MM}}_k \times {\text{MM}}_{k-1}
 \ .
\end{equation}
Similar relations between 'generalized minimal models' and
Virasoro minimal models were first discussed in \cite{Crnkovic:1989gy},
\cite{Crnkovic:1989ug} and later (it seems independently) by \cite{Lashkevich:1992sb},\cite{Lashkevich:1993fb}.
More recently, results for the 4D gauge theories \cite{Bonelli:2011kv}
inspired  Wyllard \cite{Wyllard:2011mn}
to propose an extension to cosets
of the type $(SU(N)_\kappa \times SU(N)_p )/SU(N)_{\kappa+p}$
where  $\kappa$ is a free parameter. Soon after this paper
had appeared, the case of $N=2, p=2$ was considered in more
detail by Belavin et al. \cite{Belavin:2011sw}.

Let us now describe the content of this work in more detail.
We shall begin with a brief review of Liouville field theory
and its \cN=1 supersymmetric version in the next section. Both
theories were solved long ago, see section 2 for references to
the original literature. Then we turn to imaginary Liouville
theory. As mentioned before, this model is very poorly
understood. After a few historical comments we shall
describe the 3-point functions that were proposed by
Zamolodchikov in \cite{Zamolodchikov:2005fy}. Our new
results are formulated and analyzed in section 4. There
we shall spell out a precise relation between an infinite
tower of fields in \cN=1 Liouville field theory and double
Liouville theory. This relation will be checked through
extensive comparison of 3-point functions on both sides
of the correspondence. Applications and extensions of our
results are sketched in the concluding section.

\section{Review of Liouville field theory}
\def\cB{\mathcal{B}}

In this section we simply review some basic facts about
Liouville field theory and its \cN=1 supersymmetric cousin. Most
importantly, we shall discuss the spectrum of primary fields along
with their 2- and 3-point functions. For a more details see the
reviews \cite{Teschner:2001rv,Schomerus:2005aq,Nakayama:2004vk}.

\subsection{Bosonic Liouville field theory}
\def\ta{\tilde \alpha}

Liouville field theory involves a single scalar field with
an exponential interaction term. On a 2-dimensional world-sheet
with metric $\c^{ab}$ and curvature $R$, the action of Liouville
theory takes the form
\be \label{actLiouv}
   S_L[X] \ = \ \frac{1}{4\pi}\int_\Sigma d^2 \sigma \sqrt \c
     \left( \c^{ab} \pl_a X \pl_b X +
      R Q X
    + 4 \pi \mu_L e^{2bX} \right)
\ee
where $\mu_L$ and $b$ are two (real) parameters of the model.
The second term in this action describes the background charge
of a linear dilaton.
The value of the constant $Q$ must be adjusted to the choice
of $b$ in order for $S_L$ to define a conformal quantum field
theory. We shall state the relation in a moment.

Liouville theory should be considered as a marginal deformation
of the free linear dilaton theory. The Virasoro field of a linear
dilaton theory is given by the familiar expression
$$ T(z) \ = \ - (\partial X)^2 + Q \pl^2 X \ \ . $$
The modes of this field form a Virasoro algebra with
central charge $c_L = 1 + 6 Q^2$. Furthermore, the usual
closed string vertex operators
\begin{equation}
 V_\a(z) \ =\  :\exp 2\a X(z,\bar z): \ \ \ \ \mbox{ have } \ \
 \Lcd_\a \, = \, \a (Q-\a)\,  = \, \bLcd_\a\ \ .
\label{Lcd} \end{equation}
Here and in the following we shall not explicitly display the
dependence of our vertex operators on the complex conjugate
$\bar z$ of the world-sheet coordinate $z$. Note the conformal
weights $\Lcd,\bLcd$ are real if $\a$ is of the form $\a = Q/2
+ iP$.  In order for the exponential potential in the
Liouville action to be marginal, i.e.\ $(\Lcd_b,\bLcd_b) = (1,1)$, we
must now also adjust the parameter $Q$ to the choice of $b$ in
such a way that
$$ Q = b + b^{-1} \ \ . $$
Weyl invariance of the classical action $S_L$ leads to the relation
$Q_{c} = b^{-1}$ and the additional shift by $b$ may be considered
 as a quantum correction of the classical relation. The extra term,
which certainly becomes small in the semi-classical limit $b
\rightarrow 0$, renders $Q = Q_c + b$ (and hence the central charge)
invariant under the replacement $ b \rightarrow b^{-1}$.

The solution of Liouville field theory is completely described
by the 2- and 3-point functions of the model. The vertex operators
$V_\a$ are introduced such that their 2-point function is canonically
normalized, i.e.
\begin{equation}
\langle V_{\a_2}(z_2) V_{\a_1} (z_1) \rangle
= |z_{12}|^{-4\Lcd_{\alpha_1}} 2 \pi \left( \delta(\a_1+\a_2-Q) +
D_L(\alpha_1) \delta(\a_2-\a_1)\right)
\end{equation}
where
\begin{equation} \label{L2pt}
D_L(\alpha) =  \left(\pi \mu_L \gamma(b^2) \right)^{(Q-2\alpha) \over b}
 { \gamma(2 \alpha b - b^2) \over b^2 \, \gamma(2 - 2\alpha b^{-1}+ b^{-2})}
\end{equation}
Here and throughout the main text we use $\gamma(x) = \Gamma (x)/
\Gamma(1-x)$. In order to spell out the 3-point functions we need to introduce
Barnes' double $\Gamma$-function $\Gamma_b(y)$. It may be defined
through  the following integral representation,
\begin{equation}\label{BGamma}
 \ln \Gamma_b(y) \ = \ \int_0^\infty \frac{d\t}{\t}
  \left[ \frac{e^{-y\t} - e^{- Q \t/2}}
              {(1-e^{-b \t}) (1- e^{-\t/b})}
         - \frac{\left(\frac{Q}{2} - y\right)^2}{2} e^{-\t}
           - \frac{\frac{Q}{2} -y }{\t}\right]
\end{equation}
for all $b \in \QR$. The integral exists when $0 < {\rm Re}(y)$ and
it defines an analytic function which may be extended onto the
entire complex $y$-plane. Under shifts by $b^{\pm 1}$, the function
$\Gamma_b$ behaves according to
\begin{equation}\label{BGshift}
\Gamma_b(y+b) \ = \  \sqrt{2\pi} \, \frac{b^{by-\frac12}}{\Gamma(by)}\,
                      \Gamma_b(y) \ \ , \ \
\Gamma_b(y+b^{-1}) \ = \ \sqrt{2\pi} \, \frac{b^{-\frac{y}{b}+\frac12}}
      {\Gamma(b^{-1}y)}\, \Gamma_b(y) \ \ .
\end{equation}
These shift equations let $\Gamma_b$ appear as an interesting
generalization of the usual $\Gamma$ function which may also be
characterized through its behavior under shifts of the argument.
But in contrast to the ordinary $\Gamma$ function, Barnes' double
$\Gamma$ function satisfies two such equations which are independent
if $b$ is not rational. We furthermore deduce from eqs.\
(\ref{BGshift}) that $\Gamma_b$ has poles at
\begin{equation} \label{poles}
 y_{n,m} \ = \ - n b - m b^{-1} \ \ \ \mbox{ for } \ \ \
   n,m \ = \ 0,1,2, \dots \ \ .
\end{equation}
From Branes' double Gamma function one may construct the
following basic building block of the 3-point function,
\begin{equation} \label{ZY}
 \Y_b (\a) \ := \ \Gamma_2(\a|b,b^{-1})^{-1}\,
                    \Gamma_2(Q-\a|b,b^{-1})^{-1} \ \ .
\end{equation}
The properties of the double $\Gamma$-function imply that $\Y$
possesses the following integral representation
\begin{equation}\label{ZYpsilon}
\ln \Y_b(y) \ = \ \int_0^\infty \frac{dt}{t}
   \left[ \left(\frac{Q}{2}-y\right)^2
     e^{-t} - \frac{\sinh^2\left(\frac{Q}{2} -y\right) \frac{t}{2}}
        {\sinh\frac{bt}{2}\, \sinh \frac{t}{2b}}\right]\ \ .
\end{equation}
Moreover, we deduce from the two shift properties (\ref{BGshift})
of the double $\Gamma$-function that
\begin{equation}\label{ZYshift}
   \Y_b(y+b) \ = \ \gamma(by) \, b^{1-2by}\, \Y_b(y) \ \ , \ \
   \Y_b(y+b^{-1}) \ = \  \gamma(b^{-1}y) \, b^{-1+2b^{-1}y}\, \Y_b(y)
\ \ .
\end{equation}
Note that the second equation can be obtained from the first with the help
of the self-duality property $\Y_b(y) = \Y_{b^{-1}}(y)$.
\smallskip

After this preparation it is easy to spell out the 3-point function
of primary fields in Liouville field theory \cite{Dorn:1994xn,Zamolodchikov:1995aa},
\begin{equation}\label{Cc>}
\langle V_{\a_3}(z_3)  V_{\a_2}(z_2)
V_{\a_1} (z_1) \rangle = \frac{C_L(\a_3,\a_2,\a_1|b)}
{|z_{12}|^{2h_{12}} |z_{13}|^{2h_{13}} |z_{23}|^{2h_{23}}}
\end{equation}
with $h_{12} = h_{\a_1}+ h_{\a_2}-h_{\a_3}$ etc. and coupling constants
$C_L$ of the form
\begin{equation}\label{Lstr}
C_L(\a_3,\a_2,\a_1|b) = \left[\pi \mu_L \gamma(b^2) b^{2-2b^2}
\right]^{\frac{Q-\talpha}{b}} \!\! \frac{\Upsilon^0_b \, 
\Upsilon_{b}\left( 2 \alpha_1\right) \Upsilon_{b}\left( 2 \alpha_2\right) \Upsilon_{b}\left( 2 \alpha_3\right) }{ \Upsilon_{b}\left( \talpha_{123} - Q\right)\,
\Upsilon_{b}\left( \talpha_{12}\right)\, \Upsilon_{b}\left( \talpha_{13}\right)\, \Upsilon_{b}\left( \talpha_{23}\right) }\ .
\end{equation}
Here and in the following, the contant $\Upsilon^0_b$ is given by
$\Upsilon^0_b = \Upsilon'_{b}\left( 0\right)$. Furthermore, the parameters
$\talpha_{123}$ and $\talpha_{ij}$ are certain linear combinations of $\a_j$,
$$ \talpha_{123} = \a_1+\a_2+\a_3 \ , \ \talpha_{12} = \a_1 + \a_2 - \a_3 \quad \mbox{etc.}$$
The solution (\ref{Lstr}) was first proposed by H.\ Dorn
and H.J.\ Otto \cite{Dorn:1994xn} and by A.\ and Al.\ Zamolodchikov
\cite{Zamolodchikov:1995aa}, based on extensive earlier work by many
authors (see e.g.\ the reviews \cite{Seiberg:1990eb,Teschner:2001rv,
Schomerus:2005aq} for references). Full crossing symmetry of the
conjectured 3-point function was established much later in two steps
by Ponsot and Teschner \cite{Ponsot:1999uf} and by Teschner
\cite{Teschner:2001rv,Teschner:2003en}. The proof of consistency
of the DOZZ structure constants for Liouville field theory was
completed recently by establishing modular invariance of 1-point
functions on a torus \cite{Hadasz:2009sw}.

\subsection{\cN=1 Liouville field theory}

\cN=1 supersymmetric Liouvilel field involves one real superfield that
contains a real bosonic scalar $\varphi$, the two components $\psi$ and
$\bar \psi$ of a Majorana fermion and an auxiliary field $F$. After
integrating out the latter and fixing the world-sheet metric, the 
action of \cN=1 super Liouville field theory takes the form
\begin{align}
   S_{SL}[\varphi,\psi]
   \ = \ \frac{1}{2\pi} \int d^2 z \left[  \partial \varphi \bar
   \partial \varphi
  + \psi \bar \partial \psi + \bar \psi \partial \bar \psi \right] +
  2 i \mu b^2 \int d^2 z \psi \bar \psi e^{ b \varphi}  ~,
\end{align}
The background charge for the boson $\varphi$ is related to the 
parameter $b$ by $Q = b + 1/b$. As in the case of bosonic Liouville field theory,
the supersymmetric cousin is obtained by perturbing a free field theory,
namely the product of a linear dilaton with a 2-dimensional Ising model.
The spectrum of the Ising model contains six conformal blocks including
the identity field, the two components $\psi$ and $\bar \psi$ of the
fermion and the energy density $\psi \bar \psi$, which are all part of
the Neveu-Schwarz (NS) sector. In addition, there are two blocks in the
Ramond (R) sector. These are generated from the spin field $\varsigma^+ =
\sigma$ and the so-called disorder field $\varsigma^- = \mu$. After
multiplication with the linear dilator, the model contains an \cN=1
super-conformal symmetry with central charge $c_{SL} = \frac{3}{2}(1 + 2
Q^2)$. The holomorphic half of this symmetry is generated by modes of
the following fields
\begin{equation}
T(z) = - \frac12\left((\partial \varphi)^2 - Q \partial^2 \varphi +
\psi \partial \psi\right)
\quad , \quad G(z) = -i( \psi \partial \varphi - Q \partial \psi ) \ .
\label{N1SCA}
\end{equation}
Anti-holomorphic fields can be constructed similarly. The
interacting theory has been solved soon after the DOZZ proposal
had been put out, see \cite{Rashkov:1996jx,Poghosian:1996dw}.
Vertex operators in the NS sector are super-descendents of
\begin{equation}
\phi_\a (z) = : \exp \a \varphi(z,\bar z) : \quad \mbox{ with } \quad \D_\alpha
= \alpha(Q-\alpha)/2
= \bar \D_\alpha  \label{NSvert}
\end{equation}
The 2-point function of these NS primary fields takes the form
\begin{align}\label{L2ptNS}
 \langle \phi_{\alpha_2} (z_2) \phi_{\alpha_1} (z_1 ) \rangle
\  =\  |z_{12}|^{-4 \Delta_{\alpha_1}}
  2 \pi \left[ \delta (\alpha_1 + \alpha_2 - Q)
+ \delta (\alpha_2 - \alpha_1 ) D_{NS}(\alpha_1) \right] ~,
\end{align}
with
\begin{align}
 D_{NS} (\alpha) \ = \ - \left(\mu \pi \gamma
    (\tfrac{bQ}{2}) \right)^{\frac{Q-2\alpha}{b}}
  \frac{\Gamma ( b (\alpha-\frac{Q}{2}) )
        \Gamma ( \frac{1}{b} (\alpha-\frac{Q}{2}) ) }
       {\Gamma (- b (\alpha-\frac{Q}{2}) )
        \Gamma ( - \frac{1}{b} (\alpha-\frac{Q}{2}) ) } ~.
\end{align}
Whereas the first term in eq.\ \eqref{L2ptNS} is fixed by normalization,
the second term involving $D_{NS}$ contains dynamical information
on the phase shift of tachyonic modes upon reflection off the Liouville
wall.
\smallskip

To spell out the  3-point functions of the model we need to build
two new special functions from the $\Upsilon$-function we introduced in the
previous subsection, see eq.\ \eqref{ZYpsilon}. These are given by
\begin{align}
 \Upsilon^{\text{NS}}_b (x) &= \Upsilon_b ( \tfrac{x}{2})
   \Upsilon_b (\tfrac{x+Q}{2}) ~,
 &\Upsilon^{\text{R}}_b (x) &= \Upsilon_b (\tfrac{x+b}{2})
   \Upsilon_b (\tfrac{x+b^{-1}}{2}) ~.
\end{align}
Properties of these new functions can easily be derived from the properties
of $\Upsilon_b$ we listed above. In particular, we note that the functions
$\Upsilon^{\text NS}_b$ and $\Upsilon^{\text R}_b$ possess the following
behavior under shifts of their argument,
\begin{align}
 \Upsilon^{\text{NS}}_b (x+b) &= b^{-b x} \gamma ( \tfrac12 + \tfrac{bx}{2})
 \Upsilon^{\text{R}}_b(x) ~,
 &\Upsilon^{\text{R}}_b (x+b) &= b^{1-b x} \gamma (\tfrac{bx}{2})
 \Upsilon^{\text{NS}}_b (x)~, \\[2mm]
  \Upsilon^{\text{NS}}_b (x+ \tfrac{1}{b})
  &= b^{\frac{x}{b}} \gamma (\tfrac12 + \tfrac{x}{2b})
 \Upsilon^{\text{R}}_b (x)~,
 &\Upsilon^{\text{R}}_b (x+ \tfrac{1}{b}) &= b^{-1+ \frac{x}{b} }
 \gamma (\tfrac{x}{2b})
 \Upsilon^{\text{NS}}_b (x) ~.
 \label{uprel}
\end{align}
The functions $\Upsilon_b^{\rm NS}, \Upsilon^{\text{R}}_b$ suffice to state the 3-point structure constants of the NS sector,
\begin{equation}\label{NS3p}
\langle \phi_{\a_3}(z_3)  \phi_{\a_2}(z_2)
\phi_{\a_1} (z_1) \rangle = \frac{C_{NS}(\a_3,\a_2,\a_1|b)}
{|z_{12}|^{2h_{12}} |z_{13}|^{2h_{13}} |z_{23}|^{2h_{23}}}
\end{equation}
\begin{equation} \label{NS3p2}
\langle \phi_{\a_3}(z_3) \tilde \phi_{\a_2}(z_2) \phi_{\a_1}(z_1)
\rangle = \frac{\tilde C_{NS}(\a_3,\a_2,\a_1|b)}
{|z_{12}|^{2h_{12}+1} |z_{13}|^{2h_{13}-1} |z_{23}|^{2h_{23}+1}}
\end{equation}
where $\tilde \phi_\a = \{ G_{-\frac12},[ \bar G_{-\frac12} ,\phi_\a]\}$, and
\begin{eqnarray}
C_{NS}(\alpha_3,\alpha_2,\alpha_1|b)&=&\frac12
\left[\frac{\pi \mu}{2b^{b^2-1}} \gamma\left(\frac{Qb}{2}\right)
\right]^{Q-\talpha_{123} \over b}\!\!\!\!\!\!
\frac{\Upsilon^0_b \,\Upsilon_b^{\rm NS}(2\alpha_1) \Upsilon_b^{\rm NS}(2\alpha_2) \Upsilon_b^{\rm NS}(2\alpha_3)}{
    \Upsilon_b^{\rm NS}(\talpha_{123}-Q) \Upsilon_b^{\rm NS}(\talpha_{12})\Upsilon_b^{\rm NS}(\talpha_{23})\Upsilon_b^{\rm NS}(\talpha_{13})}\nonumber\\[2mm]
& &  \label{SLstr}
\\[-2mm]
\tilde C_{NS}(\alpha_3, \alpha_2, \alpha_1|b)&=& i
\left[ \frac{\pi \mu}{2b^{b^2-1}} \gamma\left(\frac{Qb}{2}\right)\right]^{Q-\talpha_{123}
\over b} \!\!\!\!\!\!
\frac{\Upsilon^0_b \,\Upsilon_b^{\rm NS}(2\alpha_1) \Upsilon_b^{\rm NS}(2\alpha_2) \Upsilon_b^{\rm NS}(2\alpha_3)}{
    \Upsilon_b^{\rm R}(\talpha_{123}-Q) \Upsilon_b^{\rm R}(\talpha_{12})\Upsilon_b^{\rm R}(\talpha_{23})\Upsilon_b^{\rm R}(\talpha_{13})} \nonumber
\end{eqnarray}
Any 3-point function of descendent fields can be written  in terms of the correlator (\ref{NS3p}) or (\ref{NS3p2})
and 3-point blocks which are completely determined by the super-conformal Ward identities (see e.g. \cite{Hadasz:2006qb}).

Let us now turn to the R sector of the model. As we recalled before,
the 2-dimensional Ising model possesses two local fields of conformal
weight $\Delta = 1/16 = \bar \Delta$ which we denoted by  $\varsigma^+ =
\sigma$ and $\varsigma^- = \mu $, see chapter 12 of \cite{DiFrancesco:1997nk}
for more details. Using these spin fields, we can define the following
two vertex operators in the R sector of \cN=1 Liouville theory
\begin{align}
\Sigma^{\pm}_{\alpha}(z) &=\ \varsigma^{\pm}(z,\bar z)
\ :e^{ \alpha \varphi(z,\bar z)}: \quad \mbox{with}
\quad \Delta_\a^{\rm R} = \frac12\a(Q-\a) + \frac{1}{16} = \bar \Delta^{\rm R}_\a
\ . \label{spin}
\end{align}
Our conventions are the same as in \cite{Poghosian:1996dw, Hadasz:2008dt}
and they imply
\begin{eqnarray}\label{Rfields}
G_0 \Sigma^\pm_{\a}(z) = i \beta e^{\mp i \frac{\pi}{4} } \Sigma^\mp_{\a}(z) ,
\quad \bar G_0 \Sigma^\pm_{\a}(z) = -i \beta e^{\pm i \frac{\pi}{4} } \Sigma^\mp_{\a}(z) ,  \qquad \beta = \frac{1}{\sqrt2} \left(\frac{Q}{2} - \a \right)\, .
\end{eqnarray}
The 2-point functions of the vertex operators $\Sigma^{\epsilon}_{\alpha}$
possess the following form
\begin{align}
&\langle \Sigma^{\pm}_{\alpha_2} (z_2) \Sigma^{\pm}_{\alpha_1} (z_1) \rangle
\ = \ |z_{12}|^{- 4 \Delta_{\alpha_1} - \frac14} 2 \pi
 \left[\delta (\alpha_1 + \alpha_2 - Q) \pm \delta (\alpha_2 - \alpha_1 )
  D_R  (\alpha_1) \right] ~
 \label{L2ptR}
\end{align}
with a reflection coefficient given by
\begin{align}
 D_R (\alpha )\  = \
 \left(\mu \pi \gamma (\tfrac{bQ}{2}) \right)^{\frac{Q-2\alpha}{b}}
  \frac{\Gamma (\frac12 + b (\alpha-\frac{Q}{2}) )
        \Gamma (\frac12 + \frac{1}{b} (\alpha-\frac{Q}{2}) ) }
       {\Gamma (\frac12 - b (\alpha-\frac{Q}{2}) )
        \Gamma (\frac12 - \frac{1}{b} (\alpha-\frac{Q}{2}) ) } ~.
        \label{refR}
\end{align}
Let us also provide explicit expressions for the 3-point functions
involving two RR fields. These were determined in \cite{Rashkov:1996jx,
Poghosian:1996dw,Fukuda:2002bv} and we shall simply quote the results
along with all the necessary notations,
\begin{align}\label{SLstrR}
 \langle  \phi_{\a_3} (z_3) \Sigma_{\alpha_2}^{\pm} (z_2)
 \Sigma_{\alpha_1}^{\pm} (z_1)
 \rangle
 \ = \
    \frac{C^{\pm}_R (\alpha_3 ; \alpha_2 , \alpha_1|b)}{|z_{12}|^{2\Delta_{12}+\frac14}
   |z_{23}|^{2\Delta_{23}}
   |z_{13}|^{2\Delta_{13}}} ~.
\end{align}
The structure constants $C^{\pm}_R$ are constructed from the special
functions $\Upsilon^{\rm NS}$ and $\Upsilon^{\rm R}$ as follows,
\begin{eqnarray}\label{N1LstrR}
\nonumber
 C^{\pm}_R (\alpha_3; \alpha_2, \alpha_1|b )
  &=&\frac12 \left[ \frac{\mu \pi}{2} \gamma (\tfrac{bQ}{2} )  b^{1-b^2}
    \right]^{\frac{Q- \talpha_{123}}{b}}
    \frac{\Upsilon^{0}_b \, \Upsilon^{\text{R}}_b(2 \alpha_1)
     \Upsilon^{\text{R} }_b (2 \alpha_2 ) \Upsilon^{\text{NS}}_b (2 \alpha_3) }
     { \Upsilon^{\text{R}}_b( \talpha_{123} - Q)
     \Upsilon^{\text{R} }_b (\talpha_{12})
     \Upsilon^{\text{NS}}_b (\talpha_{23})
     \Upsilon^{\text{NS}}_b (\talpha_{13}) } \\[2pt]
     \\[-2pt]
     \nonumber
 &\pm&\!\!\!\frac12 \left[ \frac{\mu\pi}{2} \gamma ( \tfrac{bQ}{2} )
     b^{1-b^2} \right]^{\frac{Q- \talpha_{123}}{b}}
    \frac{\Upsilon ^{0}_b \, \Upsilon^{\text{R}}_b(2 \alpha_1)
     \Upsilon^{\text{R} }_b (2 \alpha_2 ) \Upsilon^{\text{NS}}_b (2 \alpha_3) }
     { \Upsilon^{\text{NS}}_b( \talpha_{123} - Q)
     \Upsilon^{\text{NS} }_b (\talpha_{12})
     \Upsilon^{\text{R}}_b (\talpha_{23})
     \Upsilon^{\text{R}}_b (\talpha_{13}) } ~.
\end{eqnarray}
Crossing symmetry of 4-point functions  in the NS sector of \cN=1 Liouville theory with  structure constants (\ref{SLstr}) and (\ref{N1LstrR}) was
first checked numerically, see \cite{Belavin:2007gz,Belavin:2007eq}, and later
proved analytically in \cite{Hadasz:2007wi,Chorazkiewicz:2008es} using braiding
and  fusion properties of the 4-point blocks. In the case of 4-point functions
containing R fields, crossing symmetry of \cN=1 Liouville theory was verified numerically in \cite{Suchanek:2010kq}. The first step
necessary for an analytical proof was presented in \cite{Chorazkiewicz:2011zd}
where braiding  properties of  the 4-point blocks were derived.

\section{Imaginary Liouville theory}

Before we can state the main results of this work, we need one more
ingredient, namely a version of Liouville field theory with central
charge $c \leq 1$. In contrast to the models we described in the
previous section, the status of the theory we are about to discuss
is less clear. In particular, the issue of crossing symmetry has
not been settled. We shall begin our exposition with a few historical
comments in the first subsection. Then we continue by listing the
proposed structure constants without much further discussion.

\subsection{Some comments on history}

In usual Liouville theory, the parameter $b$ is taken to
be real so that the corresponding central charge $c \geq 25$.
The explicit expressions for 2- and 3-point functions admit
analytic continuation to complex values of $b$ with a
non-vanishing real part. Formally, the central charge
takes values $1 < c$ in this regime. Purely imaginary
values of $b$ have been a subject of several previous
studies mostly because such values are relevant for
time-like Liouville field theory and tachyon condensation
in string theory, see e.g.\ \cite{Gutperle:2003xf,Strominger:2003fn,
Schomerus:2003vv,Fredenhagen:2004cj,McElgin:2007ak,Harlow:2011ny,
Giribet:2011zx} and further references in the more recent papers.

At least for $b=i$, it is possible to define the theory
by taking a limit starting with $b = \epsilon + i$. The
resulting theory has central charge $c=1$ and it agrees
with a certain limit of unitary minimal models. This
limit was shown to satisfy crossing symmetry
\cite{Runkel:2001ng}. It is likely that similar limits can
be taken for other purely imaginary values of $i$. But even
if such limits describe consistent local quantum theories,
they would at most be defined for a discrete set of $b$-parameters.

There is an alternative approach to defining Liouville
theory for imaginary $b$, i.e.\ for $c \leq 1$. In order
to describe how this works, let us recall a few facts
about the usual construction of the 3-point couplings
in Liouville field theory. The main idea is to evaluate
crossing symmetry for 4-point functions with three
physical and one degenerate field insertions. The operator
product of a physical with a degenerate field involves a
finite set of terms whose coefficients can be computed
in free field theory. More precisely, if we take the
degenerate field to be $V_{-b^{\pm 1}/2}$, then the
4-point function must satisfy a second order differential
equation and hence only two terms can possibly arise on
the left hand side of the operator product, e.g.\
\begin{equation} \label{degOPE}
 V_\a(w,\bw) \, V_{-b/2}(z,\bz)  \ = \
   \sum_{\pm} \ \frac{c^{\pm}_b(\a)}{|z-w|^{h_\pm}}
   \ V_{\a \mp b/2}(z,\bz) \ + \ \dots
\end{equation}
where $h_\pm = \mp b \a + Q (-b/2 \mp b/2)$. A similar
expansion for the second degenerate field is obtained
by replacing $b \rightarrow b^{-1}$. We can even be more
specific about the operator expansions of degenerate
fields because the coefficients $c^\pm$ may be determined
through a simple free field computation in the linear
dilaton background. One finds that
\begin{eqnarray}
 c^-_b(\a) & = & - \mu_L \int d^2z \, \langle \, V_{-b/2}(0,0)
 \, V_{\a}(1,1)\,  V_{b}(z,\bz)\, V_{Q-b/2-\a}
       (\infty,\infty) \, \rangle_{{\rm LD}} \nn \\[2mm]
& = & - \mu_L \pi \ \frac{\c(1+b^2) \, \c(1-2b\a)}{\c(2+b^2 - 2b\a)}
\ \    \label{cm}
\end{eqnarray}
see \cite{Schomerus:2005aq} for more details. The result in the
second line is obtained using the explicit integral formulas
that were derived by Dotsenko and Fateev. The corresponding field is
then degenerate and it possesses an operator product consisting of
two terms only. {\em Teschner's trick} converts the crossing symmetry
condition into a much simpler algebraic condition. Moreover,
since we have already computed the coefficients of operator
products with degenerate fields, the crossing symmetry equation
is in fact linear in the unknown generic 3-point couplings. One
component of these conditions for the degenerate field
$V_{-b/2}$ reads as follows
\begin{eqnarray} \label{spcross}
 0 & = & C_L(\a_1 + \frac{b}{2},\a_3,\a_4)\, c^-_b (\a_1) \cP^{--}_{+-}
    + C_L(\a_1 - \frac{b}{2},\a_3,\a_4) \, c^+_b (\a_1) \cP^{++}_{+-}
   \ \ , \\[2mm] \nonumber
\mbox{where} & & \ \ \cP^{\pm\pm}_{+-} \ = \
   \Fus{\a_1\mp b/2,}{\a_3 - b/2}{- b/2\, }{\a_3}{\ \ \a_1\ }{\a_4} \
  \Fus{\a_1\mp b/2,}{\a_3 + b/2}{-b/2\, }{\a_3}{\ \ \a_1\ }{\a_4}\ \ .
\end{eqnarray}
Note that the combination on the right hand side must vanish
because in a consistent model, the off-diagonal bulk mode
$(\a_4-b/2,\a_4+b/2)$ does not exist and hence it cannot
propagate in the intermediate channel. The required special entries
of the Fusing matrix can be expressed through a combination of
$\Gamma$ functions. Once the expressions for $c^\pm$ and $\cP$
are inserted (note that they only involve $\Gamma$ functions),
the crossing symmetry condition may be written as follows,
\be
\label{3ptse}
\frac{C_L(\a_1 + b,\a_2,\a_3)}{C_L(\a_1,\a_2,\a_3)} \ = \ -
  \frac{\c(b(2\a_1 +b))\c(2b\a_1)}{\pi \mu_L \c(1+b^2)
  \c(b(\talpha_{123}-Q))}
    \frac{\c(b(\talpha_{23}-b))}{\c(b\talpha_{13}) \c(b\talpha_{12})}
\ee
with $\c(x) = \Gamma(x)/ \Gamma(1-x)$, as before. The constraint
takes the form of a shift equation that describes how the coupling
changes if one of its arguments is shifted by $b$. Using the
symmetry $b \leftrightarrow b^{-1}$ we obtain a second shift
equation that encodes how the 3-point couplings behave under
shifts by $b^{-1}$. For irrational values of $b$, the two
shift equations determine the couplings completely, at least if
we require that they are analytic in the momenta. The unique
solution turns out to be analytic in $b$ as well so that it
may be extended to all real values of the parameter $b$.

We are now prepared to take a fresh look at the problem of
constructing imaginary Liouville theory. While the structure
constants \eqref{Lstr} are not analytic in $b$ so that their
extension to imaginary $b$ (or $c \leq 1$) may be ill-defined, the
coefficients of the shift equation \eqref{3ptse} involve only
$\Gamma$ functions so that a continuation to imaginary values
of $b$ is straight forward. If we postulate that the 3-point
couplings of imaginary Liouville theory are analytic in the
parameters $\alpha_i$ and exists for all $c \leq 1$, then
there is again a unique solution \cite{Zamolodchikov:2005fy}.
We shall describe this solution in the following subsection.

\subsection{Zamolodchikov's solution}

Imaginary Liouville theory may be thought of as a model whose
action is formally given by
\be \label{actILiouv}
   S_{\mathcal L}[\hat X] \ = \ \frac{1}{4\pi}\int_\Sigma d^2 \sigma \sqrt \c
     \left( - \c^{ab} \pl_a \hat X \pl_b \hat X +
      R \hat Q \hat X
    + 4 \pi \mu_\L e^{-2\hat b\hat X} \right)
\ee
One can obtain it the usual action of ordinary Liouville theory by
the formal replacements $X \rightarrow - i \hat X$, $b \rightarrow
-i \hat b$ and $Q \rightarrow i \hat Q$. Vertex operators in this
model take the form
\begin{equation}
{\mathcal V}_{\hat \a}(z) = : e^{2 \hat a \hat X(z,\bar z) }
\quad \mbox{ with } \quad \hat \Lcd_{\hat \a} =
- \hat \a (\hat Q - \hat \a)= \hat \bLcd_{\hat \a} \ .
\label{ILcd}
\end{equation}
They are obtained from the vertex operators of ordinary Liouville
theory if we replace $\a$ by $\a \rightarrow -\hat \alpha$. For
conformal invariance, the parameter $\hat Q$ must be adjusted to
the parameter $\hat b$ such that
\begin{equation}  \hat Q = \hat b^{-1} -\hat  b \ \ . \end{equation}
In terms of these parameters, the central charge of the Virasoro
algebra is now given by $c_{\mathcal L} = 1 - 6 \hat Q^2$.

As we have argued in the previous subsection, it is somewhat natural
to introduce the 3-point coupling of this imaginary Liouville theory
such that it the shift equation \eqref{3ptse} is satisfied. In terms
of the real parameters $\hat \a$ and $\hat b$, the shift equation
reads
\be
\label{IL3ptse1}
\frac{C_{\L}(\hat \a_1 - \hat b,\hat \a_2,\hat \a_3)}
{C_{\L}(\hat \a_1,\hat \a_2,\hat \a_3)} \ = \ -
  \frac{\c(\hat b(2\hat \a_1 - \hat b))\c(2\hat b\hat \a_1)}{\pi \mu_\L
  \c(1-\hat b^2)
  \c(\hat b(\hat \talpha_{123}-\hat Q))}
    \frac{\c(\hat b(\hat \talpha_{23}+\hat b))}{\c(\hat b\hat \talpha_{13})
    \c(\hat b\hat \talpha_{12})}\ .
\ee
Here we have simply carried out the substitutions we listed after
eqs.\ \eqref{actILiouv} and \eqref{ILcd}. If we shift $\hat \a_1$
by $\hat \b$ and invert the relation we obtain,
\be
\label{IL3ptse2}
\frac{C_{\L}(\hat \a_1 + \hat b,\hat \a_2,\hat \a_3)}
{C_{\L}(\hat \a_1,\hat \a_2,\hat \a_3)} \ = \
 - \frac{\pi \mu_\L
  \c(\hat b(\hat \talpha_{123}-\hat b^{-1} + 2 \hat b))}
  {\c(\hat b^2)\c(\hat b(2\hat \a_1))\c(2\hat b(\hat \a_1+\hat b))}
    \frac{\c(\hat b(\hat \talpha_{13}+\hat b))
    \c(\hat b(\hat \talpha_{12}+\hat b))}
    {\c(\hat b\hat \talpha_{23})}\ .
\ee
Note that all the factors that depend on linear combination of
the variables $\hat \a_i$ are the same as in eq.\ \eqref{3ptse},
except for a simple shift by $\hat b$. Factors depending on $\a_1$
are not universal since they are effected by the normalization of
vertex operators. Following \cite{Zamolodchikov:2005fy} we fix the
normalization such that
\begin{eqnarray*}
\langle \mcV_\halpha(z_2) \mcV_\halpha(z_1)\rangle & = &
   |z_{12}|^{-4 \Lcd_\halpha} G(\halpha) \ , \\[2mm]
   G(\halpha) &=&  \left(\pi \mu_{\L} \gamma(-\hat b^2) \right)^{2\halpha
  \over \hat b} { \gamma(2 \halpha \hat b + {\hat b^2}) \, \gamma(2-\hat b^{-2})
  \over  \gamma(2 + 2\halpha \hat b^{-1}- \hat b^{-2}) \, \gamma(\hat b^2) }\ .
\end{eqnarray*}
The expression on the left hand side is obtained from the second
term in eq.\ \eqref{L2pt} by our standard substitutions. Once this
normalization is adopted, the associated 3-point couplings take the
form
\begin{eqnarray} \label{ILstr}
C_{\L}(\halpha_3, \halpha_2, \halpha_1|\hat b) & = &  \left( \pi \mu_\L
\gamma\left(-\hat b^2\right) \right)^{\halpha_{123} \over \hat b}
\, b^{2(b+b^{-1})(\halpha_{123}
+\hat b - \frac{1}{\hat b}) }
\frac{\gamma\!\left(2- \hat b^{-2} \right)}{\gamma\!\left(\hat b^{2} \right)}\, \hat b^2\\[2mm]
& & \hspace*{-2cm} \times \frac{\Upsilon_{\hat b}\left( \halpha_{123} - \hat b^{-1}+2\hat b\right)
\Upsilon_{\hat b}\left( \halpha_{12} + \hat b \right)\Upsilon_{\hat b}\left(
\halpha_{23} + \hat b \right)\Upsilon_{\hat b}\left( \halpha_{13} + \hat b \right)}
{\Upsilon_{\hat b}^0\, \Upsilon_{\hat b}\left( 2 \halpha_1 +\hat b \right) \Upsilon_{\hat b}\left( 2 \halpha_2 +\hat b \right) \Upsilon_{\hat b}\left( 2 \halpha_3 +\hat b \right)} \ . \nonumber
\end{eqnarray}
It is easy to check that these structure constants solve the shift
equations \eqref{IL3ptse2}, though with a different $\a_1$-dependent
prefactor. This concludes our presentation of imaginary Liouville
theory.

\section{Bosonization of \cN=1 Liouville field theory}

It is well known \cite{DiFrancesco:1997nk} that a certain orbifold of
the product of two real fermions can be bosonized, i.e.\ it is
equivalent to a compactified free boson with compactification
radius $R =1$. We will now show that a similar bosonization
exists for an orbifold of the product of \cN=1 Liouville field
theory with a free fermion $\eta$. In this case, the bosonic
description involves two Liouville fields, one with real and
the other with imaginary parameter $b$. This relation was
first conjectured in \cite{Belavin:2011sw} for the Neveu-Schwarz
sector of the supersymmetric Liouville field theory. We will extend the
correspondence to the Ramond sector and perform extensive
tests for a number of local 3-point functions.

\subsection{Product of \cN=1 Liouville and a fermion}

Before we discuss the product of \cN=1 Liouville theory and a
free fermion $\eta$, let us briefly review a few things about
a product of fermions. As before, we shall denote one of our
fermions by $\psi, \bar\psi$ and the other by $\eta, \bar \eta$.
Both $\psi$ and $\eta$ are assumed to possess the same standard
operator product, i.e.
$$ \psi(z) \psi(w) \sim \frac{1}{z-w} \quad , \quad
    \eta(z) \eta(w) \sim \frac{1}{z-w} \ . $$
While the first fermion $\psi$ is assumed to be real, i.e.\
$\psi^\dagger = \psi^* = \psi$, we will modify the usual conjugation
for $\eta$ such that $\eta^\dagger = -\eta$. In this sense, the
fermion $\eta$ may be considered imaginary. Note that the usual
conjugation $\eta^* = \eta$ differs from the conjugation
$\dagger$ by a simple automorphism of the fermionic theory. In fact,
the map $\eta \rightarrow - \eta$ preserves the operator product
of the fermion $\eta$. While the algebraic properties of the two
fermions are identical, we will use a different bilinear form on
their state spaces, one that preserves $\dagger$ rather than the
usual $\ast$. As one can easily see, this form is indefinite. Our
choice will be motivated a posteriori through the relation with
double Liouville theory (see next subsection). Alternatively, one
may observe that an imaginary fermion $\eta$ emerges naturally in
the reduction from the OSP(1$|$2) WZW model to \cN=1 Liouville
field theory (see formula (2.17) of \cite{Hikida:2007sz}).

The theory of a single fermion possesses six conformal primaries,
namely the identity, the fermion fields, the energy density and
two spin fields. The latter will be denoted by $\varsigma^\pm$
and $\sigma^\pm$ for fermions $\psi$ and $\eta$, respectively.
In order to fix our conventions for $\sigma^\pm$, let us state
the analogue of the relations \eqref{Rfields}
\begin{equation}
\eta_0 \sigma^\pm = \frac{1}{\sqrt2}  \, e^{\mp i \frac{\pi}{4}}
\, \sigma^\mp, \qquad \bar \eta_0 \sigma^\pm = \frac{1}{\sqrt2}
\, e^{\pm i \frac{\pi}{4}} \, \sigma^\mp\ .
\end{equation}
Here, $\eta_0$ and $\bar \eta_0$ denote the zero modes of the
fermionic fields $\eta$ and $\bar \eta$, respectively. Due to
the conjugation rules of the fermion, i.e.\ $ \eta_{-n}^\dagger
= - \eta_{n}$ and $\bar \eta_{-n}^\dagger = - \bar \eta_{n}$,
the norms of the R fields satisfy $\langle \sigma^- | \sigma^-
\rangle = - \langle \sigma^+ | \sigma^+ \rangle.$ Hence, one
of the states $|\sigma^\pm\rangle$ has negative norm.

Coming back to the product theory between the fermion $\psi$
and $\eta$, we note that it contains a closed subset of even
local fields given by
\begin{equation}
 1\ , \ \psi\bar \psi\ , \ \psi \eta\ ,\  \psi \bar \eta\ ,\
\bar \psi \eta\ ,\ \bar \psi \bar \eta\ , \ \eta \bar \eta\ ,\
\psi\bar\psi \eta \bar \eta\ ;\
r^\pm = \frac12(\varsigma^+\sigma^+ \pm \varsigma^- \sigma^-)\ .
\label{list}
\end{equation}
The associated conformal blocks give rise to a modular invariant
partition function
\begin{equation}
 Z_{\text{\rm fermion}}(q,\bar q) = |\chi_{(0,0)}+ \chi_{(\frac12,\frac12)}|^2
 + |\chi_{(0,\frac12)}+ \chi_{(\frac12,0)}|^2  + 2  |\chi_{(\frac{1}{16},
 \frac{1}{16})}|^2
\end{equation}
where $\chi_{(h,h')} = \chi^1_h \chi^2_{h'}$ are the characters of $c=1/2$
Virasoro representations with lowest weight $h = h_\psi$ and $h' = h'_\eta$.
All the fields that are included in $Z_{\text{fermion}}$ can be bosonized
through a single bosonic field $Y$ at compactification radius $R = 1$
\cite{DiFrancesco:1997nk}. The exponential fields $\exp(ikY)$ possess a
rather simple expression in terms of
$\chi = \psi + i \eta$
along with the two spin
fields $r^\pm$ we introduced in eq.\ \eqref{list}. For $k \in \mathbb{Z}$
and  $k \geq 0$ one finds
\begin{eqnarray}
:\!e^{ikY}\!:\ = \ :\!\prod_{i=0}^{k-1} \frac{1}{k!}\, \partial^{(i)} \chi
\, \bar\partial^{(i)}\bar\chi\!:\ .
\label{fermionization1}
\end{eqnarray}
 When $k \in \mathbb{Z}+ \frac12$ we need to use the spin fields
$r^\pm$. For $k \geq 1/2$ one has
\begin{eqnarray}
:\! e^{ikY}\!: \ =\  :\!r^+\!\!\prod_{i=1}^{k-1/2} \frac{1}{k!}\, \partial^{(i)}
\chi\, \bar\partial^{(i)}\bar\chi\!: \ .
\label{fermionization2}
\end{eqnarray}

We shall now replace the first fermion by \cN=1 Liouville field theory with
a Liouville field $\varphi$ and a fermion $\psi$. The total central charge of
our product theory is
\begin{equation}\label{centralN1F}
 c = c_{SL}+ \frac12 = 2 + 3\left(b + \frac{1}{b}\right)^2 = 8 +  3 b^2 + 3 b^{-2}\ .
\end{equation}
In our construction of fields we restrict to the even ones, just as for the
free fermion model we described above. For $k = 0,1/2,1,3/2, \dots$ we set
\begin{equation}
\label{Phik}
 \Phi_\a^{(k)}(z,\bar z) =
:\exp \left({\alpha \varphi(z,\bar z) +  ikY(z,\bar z)}\right):\ .
\end{equation}
The fields $\Phi_\a^{(k)}$ differ from those introduced in \cite{Belavin:2011sw}
by their normalization (see also comments below). For negative $k = -1/2,-1,-3/2,
\dots$ we introduce $\Phi^{(k)}_\a$ through the simple prescription
\begin{equation}
 \Phi_\a^{(k)}(z,\bar z) = \Phi_{Q-\a}^{(-k)}(z,\bar z) \, =\,
:\exp \left({(Q-\alpha) \varphi(z,\bar z) -  ikY(z,\bar z)}\right):\ .
\end{equation}
Up to the normalization we mentioned before, the fields $\Phi_\a^{(-|k|)}$ also
agree with those defined in \cite{Belavin:2011sw}. The conformal weight of
$\Phi_\a^{(k)}$ is given by
$$
\Delta_\a + \frac{k^2}{2} = \frac12 \alpha(Q-\alpha) + \frac{k^2}{2} \ .
$$
We note that fields with $|k| \leq 1/2$ are primary with respect to the
product of the \cN=1 super-conformal algebra and the free fermion $\eta$.
These primary fields are given by $\Phi^{(0)}_\a = \phi_\a$ and
\begin{equation}
\Phi^{(-\frac12)}_\alpha = \left( \sigma^+ \Sigma^+_\alpha
-  \sigma^-  \Sigma^-_\alpha \right) \ , \quad  \quad
\Phi^{(\frac12)}_\alpha =\frac{1}{2i} \chi_0 \bar\chi_0\Phi^{(-\frac12)}_\alpha = \left(\sigma^+ \Sigma^+_\alpha  +
\sigma^- \Sigma^-_\alpha \right)\ .
\end{equation}
For all other values of $k$, the fields $\Phi^{(k)}_\a$ are descendent
fields. Our explicit computations below will only involve the case of
$k = \pm 1, \pm 3/2$. Using the definition \eqref{Phik} and eq.\
\eqref{N1SCA} one can rewrite the first few fields as descendents
with respect to the super-conformal algebra and the fermion $\eta$,
see also \cite{Belavin:2011sw},
\begin{eqnarray*} \label{Phi1}
|\Phi^{(\pm 1)}_\a\rangle \!\!\!&=&\! \!\! \Omega^{-2}_{\pm 1}(\a) \!
\Big[G_{-\frac12} \bar G_{-\frac12}+
 (\tfrac{Q}{2} \pm P)^2 \eta_{-\frac12} \bar \eta_{-\frac12}
 +
(\tfrac{Q}{2} \pm P) \!\big( \eta_{-\frac12} \, \bar G_{-\frac12} -  \bar \eta_{-\frac12}
\, G_{-\frac12} \big)\!\Big]\,  |\phi_\a\rangle
\\[4mm]
|\Phi^{(\pm \frac32)}_\alpha\rangle \!\!\! &=& \! \!\!   \chi_{-1} \bar\chi_{-1} |\Phi^{(\frac12)}_\alpha\rangle
  =
\Omega_{\pm \frac32}^{-2}(\a)
\Big[\tfrac{2}{P^{2}}  L_{-1}G_0 \bar L_{-1} \bar G_0 +
 2 (\tfrac{Q}{2}\pm P)^2 G_{-1}  \bar G_{-1}
\\[2mm]
 \!\!\! &+ &    \!\!\!
\sqrt2 \, \Omega_{\pm \frac32}(\a) (\tfrac{Q}{2} \pm P)(\eta_{-1} \bar G_{-1} - \bar \eta_{-1} G_{-1} ) \pm \tfrac{\sqrt2}{P} \Omega_{\pm\frac32}(\a) ( \eta_{-1} \bar L_{-1}  \bar G_0 - \bar \eta_{-1} L_{-1} G_0)
\\[2mm]
 \!\!\! &+ &  \!\!\!
 \Omega_{\pm\frac32}^2(\a) \eta_{-1} \bar \eta_{-1}
\pm  \tfrac{2}{P}( \tfrac{Q}{2}\pm P)(L_{-1}G_0 \bar G_{-1} + G_{-1} \bar L_{-1} \bar G_0) \Big] \, |\Phi^{(\frac12)}_\alpha\rangle
\end{eqnarray*}
Here we wrote equations between states rather than fields by means of the
usual state-field correspondence. The variable $P$ is related to $\alpha$
through $\alpha = Q/2 + P$. Finally, the pre-factor $\Omega_k(\alpha)$
 is given by
\begin{equation}\label{Omega}
 \Omega_k(\alpha) = n_k
\prod^{i+j=2|k|}_{\scriptsize \begin{array}{c}
 i,j = 1 , \\
 2|k| - i - j \in 2 \mathbb{N}
\end{array}}  (\mbox{\it sign(k)} (2\alpha - Q)  + i b + j b^{-1}),
\end{equation}
where {\it sign}$(k) = k/|k|$ denotes the sign of $k$ when $k \neq 0$
and we set {\it sign}$(0)$=1. The first two constants take the 
values
\begin{equation}\label{nk}
n_1 = 2^{-1}, \quad n_{\frac32} = 2^{-\frac32}.
\end{equation}
There exists a straightforward but cumbersome algorithm that 
computes the numbers $n_k$ for higher values of $k$. 
In \cite{Belavin:2011sw}, the factors
$\Omega_k$ were absorbed in the normalization of the fields
$\Phi^{(k)}_\a$. 

\subsection{Relation with double Liouville theory}

We are now prepared to state the main result of this work. It
relates the model described in the previous subsection to a
product of a Liouville field theory with $c^{(1)} \geq 25$ and
an imaginary Liouville theory with $c^{(2)} \leq 1$. We shall
often refer to this product as double Liouville theory.
According to \cite{Belavin:2011sw}, the $b$-parameters of the
two factors must be chosen as
\begin{equation} \label{b12}
 b^{(1)} = { 2b \over \sqrt{2-2b^2} }, \qquad
 \left(  \hat b^{(2)}\right)^{-1} = { 2 \over \sqrt{2-2b^2} }.
\end{equation}
So that the central charge is
$$ c = c_L^{(1)} + c_\L^{(2)} = 2 + 6\left(b^{(1)}+\frac{1}{b^{(1)}}\right)^2 - 6
\left(\hat b^{(2)} -
\frac{1}{\hat b^{(2)}}\right)^2 = 8 + 3b^2 + 3 b^{-2}\ .$$
Note that the sum of central charges agrees with the central charge
\eqref{centralN1F} of the model we discussed in the previous subsection. Moreover,
as was observed in \cite{Crnkovic:1989gy,Crnkovic:1989ug,Lashkevich:1992sb}, the
two Virasoro algebras of double Liouville theory can actually be reconstructed
from the super-conformal currents $T$ and $G$ along with the fermion $\eta$,
\begin{eqnarray}
\nonumber
 L_n^{(1)} &=& \frac{1}{1-b^2} L_n - \frac{1+2b^2}{2- 2b^2}
 \sum_{r=-\infty}^{\infty} r : \eta_{n-r} \eta_r:
+ \frac{b}{1-b^2} \sum_{r=-\infty}^{\infty}  \eta_{n-r} G_r,
\\[-4pt]
\label{twoVirasoro}
\\[-4pt]
\nonumber
 L_n^{(2)} &=& \frac{1}{1-b^{-2}} L_n - \frac{1+2b^{-2}}{2- 2b^{-2}}
\sum_{r=-\infty}^{\infty} r : \eta_{n-r} \eta_r: + \frac{b^{-1}}{1-b^{-2}}
\sum_{r=-\infty}^{\infty}  \eta_{n-r} G_r.
\end{eqnarray}
Similar formulas apply to the anti-holomorphic sector, of course.
As anticipated in the previous subsection, we now note that the
familiar relation $(L^{(i)}_n)^\dagger= L_{-n}^{(i)}$ requires
$\dagger$ to act as $\eta^\dagger_n = - \eta_{-n}$ on the modes
of the fermion $\eta$. In other words, the Virasoro modes in
double Liouville theory possess the usual conjugation rules
provided that the modes $L_n$ and $G_n$ do and we take $\eta$
to be imaginary.

Given such a close relation between their chiral algebras it seems
natural to look for relations between vertex operators. Following
\cite{Belavin:2011sw} let us introduce
\begin{equation} \label{defVb}
\Vb_\alpha^{(k)}(z,\bar z) = V_{\a^{(1)}+kb^{(1)}/2}(z,\bar z) \
\mcV_{\hat \a^{(2)} + k/2\hat b^{(2)}}(z,\bar z)
\end{equation}
where $2k$ is an integer, $\a$ is a complex parameter and we defined
\begin{equation}
\alpha^{(1)} =  { \alpha \over \sqrt{2-2b^2} }
, \qquad  \hat \alpha^{(2)} =  { b \alpha \over \sqrt{2-2b^2} }. \label{a12}
\end{equation}
The conformal dimension of the vertex operators \eqref{defVb} is
easy to compute with the help of the expressions \eqref{Lcd} and
\eqref{ILcd} for conformal weights in (imaginary) Liouville theory,
\begin{eqnarray*}
h_{(\a^{(1)}+kb^{(1)}/2)} +  \hat h_{(\hat \a^{(2)}+k/2\hat b^{(2)})}
= (\a^{(1)}+kb^{(1)}/2) (Q^{(1)} - \a^{(1)}-kb^{(1)}/2)
 \\
- (\hat \a^{(2)}+ k/2\hat b^{(2)}) (\hat Q^{(2)} - \hat \a^{(2)}-k/2\hat b^{(2)}) = \frac12 \a (Q-\a) + \frac{k^2}{2}\ .
\end{eqnarray*}
These weights agree with the weights of the fields $\Phi^{(k)}_\a$
we introduced in the previous section. Hence, with proper normalizations,
the 2-point functions of the fields $\Phi^{(k)}_\a$ and $\Vb^{(k)}_\a$
agree. In addition, it is not difficult to check that the fields $\Phi^{(k)}_\a$
are primary with respect to Virasoro algebras \eqref{twoVirasoro} of the Liouville
field theory and its imaginary cousin. Given these observations it is certainly
tempting to contemplate that the relation
\begin{equation}\label{main}
\Phi^{(k)}_\alpha(z,\bar z) =  {\mathcal N}_\alpha^{(k)}\
\Vb^{(k)}_\alpha(z,\bar z)
\end{equation}
might hold in arbitrary correlation functions. Through comparison of
3-point functions we shall provide very strong support in favor of
this proposal. These computations determine the normalization
${\mathcal N}_\alpha^{(k)}$ to take the form
\begin{eqnarray}
 {\mathcal N}^{( k)}_\alpha  &=&
 (-1)^{k} \,
 \tilde {\mathcal N}^{( k)}_\alpha,
 \end{eqnarray}
when $k \in \mathbb{N}$, i.e.\ in the NS sector of the theory, and
\begin{eqnarray}
 {\mathcal N}^{( k)}_\alpha  &=&
 2^{\frac{3}{4}} \,
 \tilde {\mathcal N}^{( k)}_\alpha,
 \end{eqnarray}
in R sector, i.e.\ when $k$ takes the values $k \in \mathbb{N} + \frac12$. The common
factor $\tilde {\mathcal N}^{( k)}_\alpha$ is given by
\begin{equation}
 \tilde {\mathcal N}^{( k)}_\alpha = \frac{\left[\pi \mu_L \gamma\left((b^{(1)})^2\right) \right]^{(\alpha^{(1)} +  \frac{k b^{(1)}}{2} ) / b^{(1)}}
		\left[\pi M \gamma\left(-(\bhtwo)^2 \right) \right]^{-(\halpha^{(2)}
    +  \frac{k }{2\bhtwo})/\bhtwo}}
	{ n_k^2 \, 2^{k^2} \left[\pi \mu \gamma({b Q \over 2}) \right]^{\alpha \over b}
     b^{- 2k} \left( {1-b^2 \over 2}\right)^{\frac12 + 2k} } .
\end{equation}
The factors $n_k$ were introduced in eq.\ (\ref{nk}), at least for some special
values of $k$. In order to check that the fields $\Phi^{(k)}_\a$ and $\Vb^{(k)}_\a$
can be identified in all correlation functions, we must verify that their
3-point functions agree,
\begin{equation} \label{3ptprop}
  \varpi(k_i) \kappa(b) \langle \Phi^{(k_3)}_{\bar \a_3} \Phi_{\a_2}^{(k_2)} \Phi_{\a_1}^{(k_1)} \rangle
  =  {\mathcal N}^{(k_3)}_{\bar \a_3} {\mathcal N}^{(k_2)}_{\a_2} {\mathcal N}^{(k_1)}_{\a_1}
  \langle \Vb^{(k_3)}_{\bar \a_3}   \ \Vb^{(k_2)}_{\a_2}  \ \Vb^{(k_1)}_{\a_1}\rangle\ \ ,
\end{equation}
at least up to some constant $\kappa(b)$ that can be absorbed through an
appropriate normalization of the vacuum state, see eq.\ \eqref{FUdef} for 
a concrete formula. The factors $\varpi$ will be
shown to satisfy $\varpi^4=1$. Given the complexity of the fields $\Phi^{(k)}$,
checking eq.\ \eqref{3ptprop} is a rather non-trivial task. We are not prepared
to establish the relation \eqref{3ptprop} for all possible 3-point functions,
but we have performed a number of highly non-trivial tests. These are
described in the next subsection.

\subsection{Comparison of 3-point functions}

Our goal is to check relation \eqref{main} in a few selected examples, involving
both NS and R sector fields and also super-descendent fields. Most computations
are somewhat lengthy but in principle straight forward to carry out.

\subsubsection{NS sector}

In our first example, we take all three fields of the \cN=1 Liouville theory to be
super-primaries in the NS sector. These are multiplied with the identity field
of the free fermion theory, i.e. we consider a 3-point correlator with
$\Phi^{(k)}_\a =\Phi^{(0)}_\a = \phi_\a$. Since we have checked already that
the conformal dimensions on both sides of the correspondence \eqref{main} match,
we shall put the fields at the points $z_3 = \infty$, $z_2 = 1$ and $z_1 = 0$
so that we can omit all dependence on world-sheet coordinates. The 3-point
function of $\Phi^{(0)}_\a$ is given by
$$
\langle \Phi^{(0)}_{\bar \a_3} \Phi_{\a_2}^{(0)} \Phi_{\a_1}^{(0)} \rangle =
C_{NS}(\a_3,\a_2,\a_1|b) $$
with $C_{NS}$ as given in equation \eqref{SLstr}.
 We will use the notation $\bar \a_i \equiv Q - \a_i $ for reflected momentum of the fields located at infinity.
  The other side of the
correspondence \eqref{main} is given by
\begin{eqnarray*}
\langle \Vb^{(0)}_{\bar \a_3}   \ \Vb^{(0)}_{\a_2}  \ \Vb^{(0)}_{\a_1}\rangle
&=& C_L(\alpha_3^{(1)}, \alpha_2^{(1)}, \alpha_1^{(1)}| b^{(1)}) \,
C_{\L}(\halpha_3^{(2)}, \halpha_2^{(2)}, \halpha_1^{(2)}| \bhtwo).
\end{eqnarray*}
The arguments $\a_i^{(\nu)}$ and $b^{(\nu)}$ that appear in the arguments
of the structure constants were introduced in eqs.\ \eqref{a12} and \eqref{b12}.
Explicit expressions for the structure constants can be found in eqs.\
\eqref{Lstr} and \eqref{ILstr}. Using the identities
\begin{equation} \label{Ups}
{ \Upsilon_{b^{(1)}}\left( \alpha^{(1)}\right) \over
\Upsilon_{\bhtwo}\left( \halpha^{(2)} + \bhtwo\right)}
= B(\alpha) \Upsilon_b^{\rm NS}(\alpha),
\end{equation}
where
\begin{equation} \nonumber
 \qquad B(\alpha) =
{\Upsilon_{b^{(1)}}^0 \over
\Upsilon_{\bhtwo}^0 \Upsilon_{b}^0 }
\ b^{b^2 \alpha(Q-\alpha) \over 2 -2b^2} \, \left( 1-b^2 \over 2\right)^{ \alpha(Q-\alpha) -2 \over 4},
\end{equation}
stated in (A.9) of \cite{Belavin:2011sw}, one can check that
$$
C_L(\alpha_3^{(1)}, \alpha_2^{(1)},
\alpha_1^{(1)}| b^{(1)}) \, C_{\L}(\halpha_3^{(2)}, \halpha_2^{(2)},
\halpha_1^{(2)}| \bhtwo)
= A_1 \, C_{NS}(\a_3,\a_2,\a_1|b)
$$
with
\bea
A_1 =  \frac{2 \left( \pi \mu_L \gamma\!\left({2b^2\over 1-b2}\right) \right)^{Q-\alpha \over 2 b}
\!\! \left( \pi M \gamma\!\left({b^2-1 \over 2}\right) \right)^{b a \over 1- b^2}\! \gamma\!
\left({- 2 b^2\over 1-b^2} \right) \gamma\! \left({b^2+1 \over 2}\right)}{\left( \left({\pi
\mu \over 2}\right)  \gamma\!\left({b^2+1 \over 2}\right) \right)^{Q-\alpha \over b}  \, \left({2 \over 1-b^2}\right)^{\frac32 }}\, .
\eea
Comparison of this $\a$-dependent factor with the product of the three
normalizations ${\mathcal N}^{(0)}_{\a_i}$ we introduced in the previous
subsection gives
$$ {\mathcal N}^{(0)}_{\bar \a_3} {\mathcal N}^{(0)}_{\a_2} {\mathcal N}^{(0)}_{\a_1} A_1
 = \kappa(b)  \ . $$
The function $\kappa(b)$ depends on the parameter $b$, but it is
 independent of the labels $\a_i$. Explicitly, it is given by
\begin{equation}\label{FUdef}
\kappa(b) =  \frac{ 2 \left[\pi \mu_L \gamma\left((b^{(1)})^2\right)
\right]^{Q^{(1)} \over b^{(1)}}
		}{ \left[\pi \mu \gamma({b Q \over 2}) \right]^{Q \over b} } \,
\gamma\!\left({- 2 b^2\over 1-b^2} \right) \gamma\!\left({b^2+1 \over 2}\right) \ .
\end{equation}
In conclusion, we have established eq.\ \eqref{3ptprop} for $k_i=0$ with
 $\varpi(0,0,0) = 1$.

Let us now proceed to the next and slightly more complicated
example of the relation \eqref{main} in which at least one of
the vertex operators involves super-descendents in the \cN=1
Liouville field. More specifically, let us insert one of the
operators $\Phi^{(\pm 1)}_\a$ along with two of the operators
$\Phi^{(0)}_\a$. Looking back at the explicit formulas we
spelled out at the end of section 4.1, we observe that
only the second  term from these expressions can contribute since
$\langle \eta \rangle = \langle \bar \eta \rangle = \langle \eta \bar
\eta\rangle = 0$. Hence we obtain
\begin{eqnarray*}
&& \hspace{-30pt}
\langle  \Phi^{(0)}_{\bar \a_3}   \ \Phi^{( 1)}_{\a_2}  \
\Phi^{(0)}_{\bar \a_1} \rangle
= \Omega^{-2}_1(\a_2) \langle \phi_{\bar \a_3} G_{-\frac12}\bar G_{-\frac12}\phi_{\a_2}
  \phi_{\a_1} \rangle =  \a_2^{-2} \,  \tilde C_{NS}(\alpha_3, \alpha_2, \alpha_1|b)
\end{eqnarray*}
where the evaluation of the correlator in the \cN=1 Liouville
theory uses the structure constants \eqref{SLstr}. On the other
side of our correspondence \eqref{main} one finds
$$
 \langle  \Vb^{(0)}_{\bar \a_3}  \ \Vb^{( 1)}_{\a_2}  \ \Vb^{(0)}_{\a_1}
 \rangle =
C_L(\alpha_3^{(1)}, \alpha_2^{(1)} + \frac{b^{(1)}}{2},
\alpha_1^{(1)}| b^{(1)}) \ C_{\L}(\halpha_3^{(2)}, \halpha_2^{(2)}
+  \frac{1}{2\bhtwo}, \halpha_1^{(2)}| \bhtwo)
$$
If we insert the explicit formulas \eqref{Lstr} and \eqref{ILstr}
for the structure constants $C_L$ and $C_\L$ along with the shift
properties \eqref{BGshift} and the identity
\begin{equation}\label{Ups2}
\Upsilon_R(\alpha) = \frac{b^{b\alpha}}{\gamma\left( \frac{b\alpha+1}{2}
 \right)} \ \Upsilon_{NS}(\alpha+ b) = B^{-1}(\alpha+b)\, \frac{ b^{b\alpha}}
 {\left( {1- b^2 \over 2}\right)^{\frac{b\alpha}{2}}}
{\Upsilon_{b^{(1)}}\left(\alpha^{(1)} + \frac{b^{(1)}}{2} \right) \over
 \Upsilon_{\bhtwo}\left(\halpha^{(2)} + \frac{1}{2\bhtwo} + \bhtwo \right) }
\end{equation}
where $B(\alpha)$ is the function defined after eq.\ (\ref{Ups}), we
can check that
\bea
C_L(\alpha_3^{(1)}, \alpha_2^{(1)}+ \frac{b^{(1)}}{2}, \alpha_1^{(1)}| b^{(1)}) \
C_\L(\halpha_3^{(2)}, \halpha_2^{(2)} +  \frac{1}{2\bhtwo}, \halpha_1^{(2)}| \bhtwo)
= A_2 \, \tilde C_{NS}(\alpha_3, \alpha_2, \alpha_1|b)
\eea{equation}
where $A_2$ are given by
\bea
A_2 = - \frac{\left( \pi \mu_L \gamma\!\left({2b^2\over 1-b2}\right) \right)^{Q-\alpha-b \over 2 b} \, \left( \pi M \gamma\!\left({b^2-1 \over 2}\right) \right)^{b a +1 \over 1- b^2}\, \gamma\!\left({- 2 b^2\over 1-b^2} \right) \gamma\!\left({b^2+1 \over 2}\right)}{i \left( \left({\pi \mu \over 2}\right) \, \gamma\!\left({b^2+1 \over 2}\right) \right)^{Q-\alpha \over b}
b^{2} \, \left( {2 \over 1-b^2}\right)^{\frac72 } \ \alpha_2^{2}
} \, .
\eea
As in the previous subsection, it is not difficult to see that the functions $A_2$
may be factorized into a product of three $\a$-dependent factors ${\mathcal N}$, i.e.
$$ {\mathcal N}^{(0)}_{\bar\a_3} {\mathcal N}^{( 1)}_{\a_2}  {\mathcal N}^{(0)}_{\a_1}
   A_2 =  i \kappa(b) \,  \a_2^{-2} \ .$$
The constant $\kappa(b)$ was introduced in eq.\ \eqref{FUdef} above.
Combining these results we conclude that eq.\ \eqref{3ptprop} also holds
for $k_1 = k_3 = 0$ and $k_2=1$ with the same constant $\kappa(b)$ as in the previous computation and $\varpi(0,1,0)= i$.

As a final check for fields from the NS sector we want to consider a
correlation function in which two fields have $k=1$,
\begin{eqnarray*}
&& \hspace{-40pt} \langle  \Phi^{(0)}_{\bar \alpha_3}  \ \Phi^{(1)}_{\alpha_2}  \ \Phi^{(1)}_{\alpha_1} \rangle = \Omega_1^{-2}(\a_1) \Omega_1^{-2}(\a_2)
\\[2mm]
\!&\times&\! \Big(
(\tfrac{Q}{2} + P_1 )^2(\tfrac{Q}{2} + P_2 )^2  \langle \phi_{\a_3} \, \eta_{-\frac12} \bar \eta_{-\frac12} \phi_{\a_2} \,
 \eta_{-\frac12} \bar \eta_{-\frac12} \phi_{\a_1} \rangle
 +
 \langle \phi_{\a_3} \, G_{-\frac12}\bar G_{-\frac12}\phi_{\a_2} \, G_{-\frac12}\bar G_{-\frac12}\phi_{\a_1} \rangle
\\[2mm]
 \!&+&\!
({\tfrac{Q}{2}} +  P_1 )(\tfrac{Q}{2} + P_2 )\left( \langle \phi_{\a_3} \, \eta_{-\frac12} \bar G_{-\frac12}\phi_{\a_2} \, \eta_{-\frac12} \bar G_{-\frac12}\phi_{\a_1} \rangle
+
\langle \phi_3 \, \bar \eta_{-\frac12}  G_{-\frac12}\phi_2 \,\bar \eta_{-\frac12}  G_{-\frac12}\phi_1 \rangle \right) \Big)
\end{eqnarray*}
It can be reduced to the basic structure constants with the help of the super-conformal
Ward identities \cite{Hadasz:2006qb}
\begin{eqnarray*}
\langle \varphi_3 \,G_{k}\varphi_2(z,\bar z) \, \varphi_1\rangle &=&
 \sum\limits_{m=0}^{k+{1\over 2}}
 \left(
\begin{array}{c}
\scriptstyle k+{1\over 2}\\[-6pt]
\scriptstyle m
\end{array}
\right)
 (-z)^{m}
\left( \langle G_{m-k}\varphi_3 \,\varphi_2(z,\bar z) \,\varphi_1 \rangle\right.
\\[1mm]
 &&\hspace{20pt} - \epsilon  \; \left.
 \langle \varphi_3 \, \varphi_2(z,\bar z) \,G_{k-m}\varphi_1 \rangle \right),
\hskip 5mm k\geqslant-\scriptstyle {1\over 2},
\\[10pt]
\langle \varphi_3 \,G_{-k}\varphi_2(z,\bar z) \,\varphi_1\rangle &=&
 \sum\limits_{m=0}^{\infty}
 \left(
\begin{array}{c}
\scriptstyle k-{3\over 2}+m\\[-6pt]
\scriptstyle m
\end{array}
\right)
z^{m}
\langle G_{k+m} \varphi_3 \,\varphi_2(z,\bar z) \,\varphi_1 \rangle
\\[1mm]
 && \hspace{-95pt} -\;
\epsilon (-1)^{k+\frac12 }
\sum\limits_{m=0}^{\infty}
 \left(
\begin{array}{c}
\scriptstyle k-{3\over 2}+m\\[-6pt]
\scriptstyle m
\end{array}
\right)
 z^{-k-m+{1\over 2}}
\langle \varphi_3 \,\varphi_2(z,\bar z) \, G_{m-{1\over 2}}\varphi_1 \rangle, \hskip 5mm k>\scriptstyle {1\over 2},
\end{eqnarray*}
\begin{eqnarray*}
\langle G_{-k} \varphi_3 \,\varphi_2(z,\bar z) \,\varphi_1 \rangle
=
\epsilon
\langle \varphi_3\,\varphi_2(z,\bar z) \,G_k\varphi_1 \rangle
+\!\!\! \sum\limits_{m=-1}^{l(k-{1\over 2})} \!\!\!
 \left( \!\!
\begin{array}{c}
\scriptstyle k+1/2\\[-6pt]
\scriptstyle m+1
\end{array} \!\!
\right)
  z^{k-{1\over 2} -m}
 \langle \varphi_3\,G_{m+{1\over 2}}\varphi_2(z,\bar z) \,\varphi_1 \rangle,
\end{eqnarray*}
where $\epsilon$ denotes the parity of the field $\varphi_2$ and $l(n) = n$
for $n + 1 \geq 0$ while $l(n) = \infty$ for $n + 1 < 0$. The result reads
\begin{eqnarray*}
 \langle  \Phi^{(0)}_{\bar \alpha_3}  \ \Phi^{(1)}_{\alpha_2}  \ \Phi^{(1)}_{\alpha_1} \rangle \!&=&\!
\Omega_1^{-2}(\a_1) \Omega_1^{-2}(\a_2)\big((\tfrac{Q}{2} +  P_1 )^2(\tfrac{Q}{2} + P_2 )^2    + (\Delta_3-\Delta_2 - \Delta_1)^2
\\[2mm]
 \!&&\! +
  2 (\tfrac{Q}{2} + P_1 )(\tfrac{Q}{2} + P_2 ) (\Delta_3-\Delta_2 - \Delta_1) \big)
   C_{NS}(\alpha_3, \alpha_2, \alpha_1|b)
\\[2mm]
 \!&=&\!
\frac{(\tfrac{Q}{2} +  P_1 + P_2 -P_3)^2 \, (\tfrac{Q}{2} + P_1 +  P_2 +P_3)^2}{4  \alpha_1^2 \, \alpha_2^2 } \,
 C_{NS}(\alpha_3, \alpha_2, \alpha_1|b) \, .
\end{eqnarray*}
On the other side we find
\begin{eqnarray*}
&& \hspace{-40pt} \langle  \Vb^{(0)}_{\bar \a_3}  \ \Vb^{(1)}_{\a_2}  \ \Vb^{(1)}_{\a_1} \rangle \\
&& =
C_L(\alpha_3^{(1)}, \alpha_2^{(1)} +  \frac{b^{(1)}}{2}, \alpha_1^{(1)} +  \frac{b^{(1)}}{2}| b^{(1)}) \
C_\L (\halpha_3^{(2)}, \halpha_2^{(2)} +  \frac{1}{2\bhtwo}, \halpha_1^{(2)} +   \frac{1}{2\bhtwo}| \bhtwo)
\end{eqnarray*}
As before, comparing structure constants and using eqs.\ (\ref{Ups}) and (\ref{Ups2}) we can check that
$$
C_L(\alpha_3^{(1)} \! , \alpha_2^{(1)} \! + \! \frac{b^{(1)}}{2} \! , \alpha_1^{(1)} \! + \! \frac{b^{(1)}}{2}| b^{(1)})
C_\L(\halpha_3^{(2)}\! , \halpha_2^{(2)} \! + \!  \frac{1}{2\bhtwo}\! , \halpha_1^{(2)} \! + \! \frac{1}{2\bhtwo}| \bhtwo)
= A_3 C_{NS}(\alpha_3, \alpha_2, \alpha_1|b),
$$
where
\begin{eqnarray*}
A_3 &=&
 \frac{\left( \pi \mu_L \gamma\!\left({2b^2\over 1-b2}\right) \right)^{Q-\alpha_{123}-2b \over 2 b} \,
\left( \pi M \gamma\!\left({b^2-1 \over 2}\right) \right)^{b \a_{123} +2 \over 1- b^2}\,
\gamma\!\left({- 2 b^2\over 1-b^2} \right) \gamma\!\left({b^2+1 \over 2}\right)}{
 \left( \left({\pi \mu \over 2}\right) \, \gamma\!\left({b^2+1 \over 2}\right) \right)^{Q-\alpha_{123} \over b} }
 \\[4pt]
& \times&
2 b^{-4} \, \left( {2 \over 1-b^2}\right)^{-\frac{11}{2} } \, (2\alpha_2)^{-2}  (2\alpha_1)^{-2} \,
(\alpha_1 + \alpha_2-\alpha_3)^2 \, (\alpha_1 + \alpha_2 + \alpha_3 -Q)^2 .
\end{eqnarray*}
Thus we have
$$ {\mathcal N}^{(0)}_{\bar \a_3} {\mathcal N}^{( 1)}_{\a_2}  {\mathcal N}^{( 1)}_{\a_1}
   A_3 =  \kappa(b)  \,
\frac{(\tfrac{Q}{2} + P_1 + P_2 -P_3)^2 \, (\tfrac{Q}{2} + P_1 + P_2 +P_3)^2}{4 \alpha_2^{2}  \, \alpha_1^{2}}.
$$
Once more we have established an instance of eq.\ \eqref{3ptprop}, this time for
$k_3 = 0$ and $k_1=k_2 = 1$. The constant factor $\kappa(b)$ is given by the same
expression as in the previous two cases and $\varpi(0,1,1)=1$.

\subsubsection{R sector}

So far we have only looked at operators $\Phi^{(k)}_\a$ with $k \in \mathbb{Z}$ that
involve fields from the NS sector of the \cN =1 Liouville field theory. The correspondence
\eqref{main} we have formulated also involves fields from the R sector. These appear for
values $k \in \mathbb{Z} + \frac12$. Actually, correlators of fields in the R sector have
been one of the crucial motivations for this work, see next section. Therefore, we would
like to perform a few tests involving $\Phi^{(k)}_\a$ with $k \in \mathbb{Z} + \frac12$.

The simplest possible 3-point function involving R sector fields is the correlation
function
\begin{equation} \label{NSRR}
\langle  \Phi^{(0)}_{\a_3}  \ \Phi^{(\pm \frac12)}_{\a_2}  \ \Phi^{(\pm \frac12)}_{\a_1}
\rangle =  \langle \phi_{\a_3} \, \sigma^+_2 \Sigma^+_{\a_2} \sigma^+_1
\Sigma^+_{\a_1} \rangle + \langle \phi_{\a_3} \, \sigma^-_2 \Sigma^-_{\a_2}
\sigma^-_1 \Sigma^-_{\a_1} \rangle
\end{equation}
involving two fields from the R sector along with one from the NS sector. The
primary fields $\Sigma^\pm_\a$ of the \cN=1 Liouville field theory are accompanied
by the spin fields $\sigma^\pm$ of the free fermion. We added a subscript to these
fields in order to keep track of the insertion points. The fields $\sigma^\pm_2$
and $\sigma^\pm_1$ are inserted at $z=1$ and $z=0$, respectively. The field
inserted at $z = \infty$ involves the identity field of the free fermion model.
Hence, for the 3-point function we consider, we only need to insert 2-point
functions of the free fermion model. In passing from the left hand side of eq.\
\eqref{NSRR} we have inserted the definition of $\Phi^{(\pm \frac12)}_\a$ and we used
that $\langle \sigma^\pm_2 \sigma^\mp_1\rangle = 0$. Assuming that $\sigma^\pm$
have been normalized, the remaining 2-point functions are
$\langle \sigma^\pm_2(1)
\sigma^\pm_1(0)\rangle = \pm 1$ so that we obtain
\begin{equation} \label{NSRR2}
\langle  \Phi^{(0)}_{\bar \a_3}  \ \Phi^{(\pm \frac12)}_{\a_2}  \ \Phi^{(\pm \frac12)}_{\a_1}
\rangle =  \langle \phi_{\bar \a_3} \, \Sigma^+_{\a_2}
\Sigma^+_{\a_1} \rangle + \langle \phi_{\bar \a_3} \, \Sigma^-_{\a_2}
\Sigma^-_{\a_1} \rangle = 2 \, C^{(+)}_{R}(\a_3;\a_2,\a_1|b) .
\end{equation}
The relevant correlation functions in \cN=1 Liouville field theories were
spelled out after eq.\ \eqref{SLstrR}. With their help we find
\begin{eqnarray*} 
\nonumber
C^{(+)}_{R}(\alpha_3; \alpha_2, \alpha_1|b)  \! &=& \! \frac12\left(
C^+_{R}(\alpha_3, \alpha_2; \alpha_1|b) + C^-_{R}(\alpha_3; \alpha_2, \alpha_1|b)\right)
\\[2mm] \nonumber
&& \hspace{-40pt} =  \frac12
\left( \frac{\pi \mu}{2} \gamma\left(\frac{Qb}{2}\right) b^{1-b^2}\right)^{Q-\talpha_{123} \over b}
\
\frac{\Upsilon^0_b \,\Upsilon_b^{\rm R}(2\alpha_1) \Upsilon_b^{\rm R}(2\alpha_2) \Upsilon_b^{\rm NS}(2\alpha_3)}{
    \Upsilon_b^{\rm R}(\talpha_{123}-Q)
	\Upsilon_b^{\rm R}(\talpha_{12})
	\Upsilon_b^{\rm NS}(\talpha_{23})
	\Upsilon_b^{\rm NS}(\talpha_{13})}
\end{eqnarray*}
Similarly, we can compute
\begin{equation}\label{C(-)}
\langle  \Phi^{(0)}_{\bar \a_3}  \ \Phi^{(\frac12)}_{\a_2}  \ \Phi^{(-\frac12)}_{\a_1}
\rangle =  2 \,  C^{(-)}_R(\alpha_3; \alpha_2, \alpha_1|b)
\end{equation}
where
\begin{eqnarray*}
 C^{(-)}_{R}(\alpha_3; \alpha_2, \alpha_1|b)
 \! &=& \! \frac12\left(
C^+_{R}(\alpha_3; \alpha_2, \alpha_1|b) - C^-_{R}(\alpha_3; \alpha_2, \alpha_1|b)\right)
\\[2mm] \nonumber
&& \hspace{-40pt} = \frac12
\left( \frac{\pi \mu}{2} \gamma\left(\frac{Qb}{2}\right) b^{1-b^2}\right)^{Q-\talpha_{123} \over b}
\frac{\Upsilon^0_b \,\Upsilon_b^{\rm R}(2\alpha_1) \Upsilon_b^{\rm R}(2\alpha_2) \Upsilon_b^{\rm NS}(2\alpha_3)}{
    \Upsilon_b^{\rm NS}(\talpha_{123}-Q)
	\Upsilon_b^{\rm NS}(\talpha_{12})
	\Upsilon_b^{\rm R}(\talpha_{23})
	\Upsilon_b^{\rm R}(\talpha_{13})}
\end{eqnarray*}
On the other hand we can compute the 3-point functions of the corresponding
fields in double Liouville theory. Using  the explicit formulae (\ref{Lstr})
and the relations (\ref{Ups}), (\ref{Ups2}) one may check that
\begin{eqnarray*}
&& \hspace{-20pt} \langle  \Vb^{(0)}_{\bar \a_3}  \ \Vb^{(\pm \frac12)}_{\a_2}  \ \Vb^{(-\frac12)}_{a_1}  \rangle
= C_L(\alpha_3^{(1)} \! , \alpha_2^{(1)} \! \pm \! \frac{b^{(1)}}{4} \! , \alpha_1^{(1)} \! - \! \frac{b^{(1)}}{4}| b^{(1)})
C_\L (\halpha_3^{(2)} \! , \halpha_2^{(2)} \! \pm \!  \frac{1}{4\bhtwo} \! , \halpha_1^{(2)} \! -  \! \frac{1}{4\bhtwo}| \bhtwo)
\\ \nonumber
&& \hspace{80pt} = A_4^\pm  C^{(\mp)}_{R}(\a_3; \a_2,\a_1| b)
\end{eqnarray*}
where
\bea
A_4^\pm \!
=  \! \frac{2\left( \pi \mu_L \gamma\!\left({2b^2\over 1-b2}\right) \right)^{Q-\alpha_{123} +b/2 \pm b/2 \over 2 b} \!
 \left( \pi M \gamma\!\left({b^2-1 \over 2}\right) \right)^{b(\a_{123}-1/2 \pm 1/2) \over 1- b^2} \!\! \!
  \gamma\!\left({- 2 b^2\over 1-b^2} \right) \! \gamma\!\left({b^2+1 \over 2}\right)
 }{\left( \left({\pi \mu \over 2}\right) \, \gamma\!\left({b^2+1 \over 2}\right) \right)^{Q-\alpha \over b}
b^{-(1 \mp 1)} \, \left( {2 \over 1-b^2}\right)^{\frac12 \pm 1} }
 \, .
\eea
As in all our previous computations, it is straight forward to show that the functions
$A_4^\pm$   may be factorized into a product of three $\a$-dependent factors
${\mathcal N}$, up to the familiar $\a$-independent term \eqref{FUdef} , i.e.
$$ {\mathcal N}^{(0)}_{\bar \a_3} {\mathcal N}^{(\pm \frac12)}_{\a_2}  {\mathcal N}^{(-\frac12)}_{\a_1}
   A^\pm_4 = 2 \kappa(b) \ . $$
Combining these results we conclude once more that eq.\ \eqref{3ptprop}
holds, with the familiar $\kappa(b)$ and a factor $\varpi(0,\pm 1/2,-1/2) = 1$.

Next let us consider two  correlation functions containing the  field $\Phi_\alpha^{(\frac{3}{2})}$. In this case in order to express the correlators
in terms of the structure constants (\ref{NSRR2}), (\ref{C(-)})  one should use
the Ward identities \cite{Friedan:1984rv, Hadasz:2008dt}
\begin{eqnarray*}
&& \hspace{-20pt}
\langle G_{-n} \varphi^{\rm R}_3 \, \varphi_2(z, \bar z) \, \varphi_1^{\rm R} \rangle
=
 \epsilon \,
\langle \varphi^{\rm R}_3 \,  \varphi_2(z, \bar z)  \, G_{n} \varphi_1^{\rm R} \rangle
+
\sum\limits_{k=-\frac12}^{\infty}
 \Big(\!\!
\begin{array}[c]{c}
\scriptstyle n+ 1/2 \\[-7pt]
\scriptstyle k+1/2
\end{array}
\!\!\Big)
  z^{n-k}
 \langle \varphi^{\rm R}_3 \,  G_{k}  \varphi_2(z, \bar z)  \,\varphi_1^{\rm R} \rangle,
\\[4pt]
&& \hspace{-20pt}
\sum_{p=0}^{\infty} \left(^{\frac12}_p \right) \ z^{\frac12 -p} \
\langle \varphi^{\rm R}_3 \,  G_{p-k}  \varphi_2(z, \bar z)  \,\varphi_1^{\rm R} \rangle
=
 \sum_{p=0}^{\infty} \left(^{ \frac12-k}_{\;\;\;p} \right) (-z)^p \
\langle  G_{p+k - \frac12}  \varphi^{\rm R}_3 \,   \varphi_2(z, \bar z)  \,\varphi_1^{\rm R} \rangle
 \\
 &&\hspace{160pt} - \epsilon \sum_{p=0}^{\infty}
 \left(^{\frac12-k}_{\;\;\;p} \right)(-z)^{ \frac12 -k-p}
 \langle \varphi^{\rm R}_3 \,   \varphi_2(z, \bar z)  \, G_{p}\varphi_1^{\rm R} \rangle \, ,
\end{eqnarray*}
where $\varphi_i^{\rm R} $ denotes a R field and $\epsilon$ is the parity of the $NS$ field. Similar Ward identities apply to the fermion $\eta$.
With the help of these identities one can see that the simplest correlator with $\Phi^{(\frac32)}$ has only a few non-vanishing terms
\begin{eqnarray*}
\langle  \Phi^{(\pm \frac{1}{2})}_{\bar \a_3}  \ \Phi^{(0)}_{\a_2}  \ \Phi^{(\frac32)}_{\a_1}
\rangle &=& \Omega_{\frac32}^{-2}(\a_1) \, \langle  \Phi^{(\pm \frac{1}{2})}_{\bar \a_3}   \Phi^{(0)}_{\a_2}  \
 \Big( 2 P_1^{-2} L_{-1}G_0 \bar L_{-1} \bar G_0 + \frac12 (Q+2P_1)^2
G_{-\frac12} \bar G_{-\frac12}
 \\[2mm] && +  P_1^{-1}(Q+ 2P_1)(L_{-1}G_0 \bar G_{-1}+ G_{-1} \bar L_{-1} \bar G_0 )
\Big) \Phi_{\a_1}^{(\frac12)} \rangle,
\end{eqnarray*}
 so that
\begin{eqnarray*}
\langle  \Phi^{(\pm \frac{1}{2})}_{\bar \a_3}   \Phi^{(0)}_{\a_2}  \Phi^{(\frac32)}_{\a_1}
\rangle \!\!\! &=& \!\!\!
{i \over \Omega_{\frac32}^{2}(\a_1)}  \left( (\Delta_3 - \Delta_2 - \Delta_1) - (Q+2P_1)(P_1+P_3)/2 \right)^2
\langle  \Phi^{(\pm \frac{1}{2})}_{\bar \a_3}   \Phi^{(0)}_{\a_2}   \Phi_{\a_1}^{(-\frac12)} \rangle
 \\[2mm]
\!\!\! & = & \!\!\!
\frac{ i ( Q + 2P_1 + 2P_2 \pm  2P_3)^2  ( Q + 2P_1 - 2P_2 \pm  2P_3)^2}{  4 (2P_1 + 2b+ b^{-1})^2  (2P_1 + b+ 2b^{-1})^2}
 \, C^{(\mp)}(\a_2; \a_3,\a_1| b)
\end{eqnarray*}
The second correlator is more complicated,
\begin{eqnarray*}
&& \hspace{-20pt} \langle  \Phi^{(\pm \frac{1}{2})}_{\bar \a_3}  \, \Phi^{(1)}_{\a_2}  \, \Phi^{(\frac32)}_{\a_1}
\rangle = \Omega_{\frac32}^{-2}(\a_1)  \Omega_{1}^{-2}(\a_2)
\\[2mm]
&&  \Big\langle   \Phi^{(\pm \frac{1}{2})}_{\bar \a_3}  \left( G_{-\frac12} \bar G_{-\frac12} + (\tfrac{Q}{2}+P_2)^2 \eta_{-\frac12} \bar \eta_{-\frac12} + (\tfrac{Q}{2} +P_2) (\eta_{-\frac12} \bar G_{-\frac12} - \bar  \eta_{-\frac12} G_{-\frac12}  \right) \! \Phi^{(0)}_{\a_2}
\\[2mm]
&&  \ \times \Big(
 2^{-1} (Q+2P_1)^2 G_{-1}  \bar G_{-1} + 2^{-\frac12} \Omega_{\frac32}(\a_1) (Q+2P_1)(\eta_{-1} \bar G_{-1} - \bar \eta_{-1} G_{-1} )
\\[2mm]
&& \
+ 2 {P_1^{-2}} L_{-1}G_0 \bar L_{-1} \bar G_0 + \Omega_{\frac32}^2(\a_1) \eta_{-1} \bar \eta_{-1} + \sqrt2 \Omega_{\frac32}(\a_1)P_1^{-1} ( \eta_{-1} \bar L_{-1}  \bar G_0 - \bar \eta_{-1} L_{-1} G_0)
\\[2mm]
&& \
+  P_1^{-1}(Q+2P_1)(L_{-1}G_0 \bar G_{-1} + S_{-1} \bar L_{-1} \bar G_0) \Big) \Phi^{(\frac12)}_{\a_1} \Bigg\rangle \ .
\end{eqnarray*}
Using the Ward identities we arrive at
\begin{eqnarray*}
&& \hspace{-10pt} \langle  \Phi^{(\pm \frac{1}{2})}_{\bar \a_3}  \ \Phi^{(1)}_{\a_2}  \ \Phi^{(\frac32)}_{\a_1}
\rangle
= \Omega_{\frac32}^{-2}(\a_1) \Omega_{1}^{-2}(\a_2)  \,\Big(
(b+2 b^{-1}+2 P_1)  \left(2 b+b^{-1}+2 P_1\right)
   \left(P_2+\tfrac{Q}{2}\right)
   \\[2mm]
&&-2 (P_1\mp P_3)
   \left(\Delta_3-\Delta_2-\Delta_1-1/2 \right)+2 (2 P_1+Q) (\Delta_3-2\Delta_2-
   \Delta_1)
   \\[2mm]
&&+2 \left(P_2+\tfrac{Q}{2}\right)
   (\Delta_3-\Delta_2-\Delta_1)-(P_1\mp P_3) (2 P_1+Q)
   \left(P_2+\tfrac{Q}{2} \right)
\Big)^2
\langle  \Phi^{(\pm \frac{1}{2})}_{\a_3}  \ \Phi^{(0)}_{\a_2}  \ \Phi^{(\frac12)}_{\a_1}
\rangle
\\[2mm]
&& = \frac{ (2P_1 + 2P_2  \pm  2P_3+3b + b^{-1})^2 (2P_1 + 2P_2  \pm 2 P_3 +b + 3b^{-1})^2 \ C^{(\pm)}(\a_2; \a_3,\a_1| b)} {8 (2P_1 + 2P_2  \mp  2P_3 +Q)^{-2} (2P_2 + b+ b^{-1})^2   (2P_1 + 2b+ b^{-1})^2  (2P_1 + b+ 2b^{-1})^2 }
\end{eqnarray*}
Within double Liouville theory we find
 \begin{eqnarray*}
\langle  \Vb^{(\pm \frac12)}_{\bar \a_3}  \ \Vb^{(0)}_{\a_2}  \ \Vb^{(\frac32)}_{\a_1}  \rangle
&=&  C_L(\alpha_3^{(1)}\! \pm \frac{b^{(1)}}{4}, \alpha_2^{(1)} \! , \alpha_1^{(1)} \! + \frac{3b^{(1)}}{4}| b^{(1)}) \,
\\[2mm]
&&
C_\L(\halpha_3^{(2)} \! \pm   \frac{1}{4\bhtwo}, \halpha_2^{(2)} \! , \halpha_1^{(2)}\! +  \frac{3}{4\bhtwo}| \bhtwo)
= A_5^{\pm} C_{R}^{(\mp)}(\a_2; \a_3,\a_1| b)
\\[6pt]
 \langle  \Vb^{(\pm \frac12)}_{\bar \a_3}  \ \Vb^{(1)}_{\a_2}  \ \Vb^{(\frac32)}_{\a_1}  \rangle
 &=& C_L(\alpha_3^{(1)} \! \pm \frac{b^{(1)}}{4}, \alpha_2^{(1)} \! + \! \frac{b^{(1)}}{2} , \alpha_1^{(1)} \! + \frac{3b^{(1)}}{4}| b^{(1)}) \,
\\[2mm]
&&\hspace*{-3mm} C_\L(\halpha_3^{(2)} \! \pm  \frac{1}{4\bhtwo}, \halpha_2^{(2)} \! + \frac{1}{2\bhtwo} , \halpha_1^{(2)}\! +  \frac{3}{4\bhtwo}| \bhtwo)
= A_6^{\pm} C_{R}^{(\pm)}(\a_2; \a_3,\a_1| b)
\end{eqnarray*}
where
\bea
A_5^{\pm}
\!\! & = & \!\!
\frac{
\left( \pi \mu_L \gamma\!\left({2b^2\over 1-b2}\right) \right)^{Q-\alpha -
 3b/2 \mp b/2 \over 2 b} \,
\left( \pi M \gamma\!\left({b^2-1 \over 2}\right) \right)^{b(a+ 3/2 \pm 1/2)\over 1- b^2}  \gamma\!\left({- 2 b^2\over 1-b^2} \right) \gamma\!\left({b^2+1 \over 2}\right)
}{
\left( \left({\pi \mu \over 2}\right) \,
\gamma\!\left({b^2+1 \over 2}\right) \right)^{Q-\alpha \over b} \,
b^{3\pm 1} \, \left( {2 \over 1-b^2}\right)^{\frac{9}{2}\pm 1  }
}
\,
\\[6pt]
&&  \hspace{-20pt}
\frac{  ( Q + 2P_1 + 2P_2 \pm  2P_3)^2  ( Q + 2P_1 - 2P_2 \pm  2P_3)^2}{ 8 \,  (2P_1 + 2b+ b^{-1})^2  (2P_1 + b+ 2b^{-1})^2}
\\[8pt]
A_6^{\pm}
 \!\! & = & \!\!
-\frac{
\left( \pi \mu_L \gamma\!\left({2b^2\over 1-b2}\right) \right)^{Q-\alpha -
 5b/2 \mp b/2 \over 2 b} \,
\left( \pi M \gamma\!\left({b^2-1 \over 2}\right) \right)^{b(a+ 5/2 \pm 1/2)\over 1- b^2}  \gamma\!\left({- 2 b^2\over 1-b^2} \right) \gamma\!\left({b^2+1 \over 2}\right)
}{
\left( \left({\pi \mu \over 2}\right) \,
\gamma\!\left({b^2+1 \over 2}\right) \right)^{Q-\alpha \over b} b^{5 \pm 1} \, \left( {2 \over 1-b^2}\right)^{\frac{13}{2}\pm 1 }
}
\\[6pt]
&& \hspace{-20pt}
\frac{(2P_1 + 2P_2  \mp  2P_3 +Q)^2 (2P_1 + 2P_2  \pm  2P_3 +3b + b^{-1})^2 (2P_1 + 2P_2  \pm 2 P_3 +b + 3b^{-1})^2 }{32 \,  (2P_2 + b+ b^{-1})^2   (2P_1 + 2b+ b^{-1})^2  (2P_1 + b+ 2b^{-1})^2 }
\eea
Comparing with the correlators from the first part of the computation in
\cN=1 Liouville theory we obtain,
\begin{eqnarray*}
&& \hspace{-10pt}
 {\mathcal N}^{(\pm \frac12)}_{\bar \a_3} {\mathcal N}^{(0)}_{\a_2}  {\mathcal N}^{(\frac32)}_{\a_1}
   A^\pm_5 =  2 \kappa(b) \, \frac{  ( Q + 2P_1 + 2P_2 \pm  2P_3)^2  ( Q + 2P_1 - 2P_2 \pm  2P_3)^2}{ 8 \,  (2P_1 + 2b+ b^{-1})^2  (2P_1 + b+ 2b^{-1})^2}\ ,
\\[4pt]
&& \hspace{-10pt}{\mathcal N}^{(\pm \frac12)}_{\bar \a_3} {\mathcal N}^{(1)}_{\a_2}  {\mathcal N}^{(\frac32)}_{\a_1}
   A^\pm_6 = 4 \kappa(b) \\[2mm]
   &&  \frac{(2P_1 + 2P_2  \mp  2P_3 +Q)^2 (2P_1 + 2P_2  \pm  2P_3 +3b + b^{-1})^2 (2P_1 + 2P_2  \pm 2 P_3 +b + 3b^{-1})^2 }{32 \,  (2P_2 + b+ b^{-1})^2   (2P_1 + 2b+ b^{-1})^2  (2P_1 + b+ 2b^{-1})^2 }
\end{eqnarray*}
so that we verified two additional cases of eq.\ \eqref{3ptprop} with
$\varpi(\pm 1/2,0,3/2)= -i$ and $\varpi(\pm 1/2,1,3/2) = 1$. This concludes
the tests of our main correspondence \eqref{main}.

\section{Outlook and Conclusions}

The main result of this work is our formula \eqref{main} that relates fields in
the product of \cN=1 Liouville field theory with a free fermion to primaries in
double Liouville field theory. We have tested this proposal through a number of
non-trivial calculations. The correspondence \eqref{main} extends related
observations in \cite{Belavin:2011sw} to the R sector. In addition,
we have been able to normalize the fields in both R and NS sector such
that the 3-point functions agree up to a simple $b$-dependent factor $\sim
\kappa$. Since this factor does not depend on the fields we insert, it can be
absorbed through a normalization of the vacuum state.

Our results may be extended in a number of different directions.
It clearly seems worthwhile to study the correspondence \eqref{main}
for correlation functions e.g.\ on discs with non-trivial boundary
conditions or higher genus surfaces. \cN =1 Liouville field theory
possesses one continuous family of boundary conditions which preserve
the \cN =1 super-conformal algebra. Though boundary conditions in
imaginary Liouville theory have not received as much attention as
the bulk model, see however \cite{Fredenhagen:2004cj}, it seems likely
that double Liouville theory admits conformal boundary conditions
that are parametrized by two continuous labels. A subset of these
boundary conditions should preserve the larger \cN=1 super-conformal
symmetry along with simple gluing conditions for the fermion $\eta$.

In an interesting recent paper \cite{Gaiotto:2012wh} Gaiotto engineers a
conformal interface between the minimal models MM$_k$ and
MM$_{k-1}$. Gaiotto's construction makes essential use of the
relation \eqref{mainrat} between the product theory and
supersymmetric minimal models. It seems likely that a similar
interface between Liouville theory and its imaginary version
also exists. Constructing this interface explicitly might be
of some interest as it could provide more insight into the
relation between standard Liouville field theory and its
imaginary cousin.

The main motivation for this work, however, came from the
results of \cite{Hikida:2007sz} which relate correlators
of the OSP(1$|$2) WZW model at level $k$ to those of
\cN=1 Liuoville theory with $b^{-2} = 2k-3$. In order to
compute N-point functions of primaries $V^\epsilon_j(\mu|z)$
in the WZW model, one needs to calculate higher correlators in a
product of \cN=1 Liouville field theory with a free
fermion. The latter involve N fields from the physical
spectrum of the supersymmetric Liouville theory along
with N-2 degenerate ones whose insertion points $y_i = y_i
(\mu_\nu)$ depend on the complex parameters $\mu_\nu$. It
turns out that all these fields must be takes from the R
sector of the model. More precisely, one finds
\begin{equation}\label{OSP}
\langle \prod_{\nu=1}^N V^{\epsilon_\nu}_{j_\nu}(\mu_\nu | z_\nu)\rangle
\sim
\delta^2(\sum_{\nu=1}^N \mu_\nu)
\langle \prod_{\nu=1}^N \frac12 \left( \Phi^{(-\frac12)}_{\alpha_\nu}
(z_\nu)  - i \epsilon_\nu  \Phi^{(\frac12)}_{\alpha_\nu} (z_\nu) \right)
\prod_{j=1}^{N-2} \Phi^{(\frac12)}_{- \frac{1}{2b}} (y_j)  \rangle\ .
\end{equation}
up to some simple factors. In the present article we have
argued that the correlation functions on the right hand
side can be calculated in double Liouville theory.
We believe that such a relation between the OSP(1$|$2) WZW model
and double Liouville theory could become a crucial ingredient
in finding a supersymmetric analogue of the celebrated FZZ-duality
between the SL(2)$/$U(1) black hole sigma model and sine-Liouville
field theory, much along the lines of \cite{Hikida:2008pe}. In this
context it is crucial to observe that, according to eq.\ \eqref{main},
the degenerate fields we have to insert at points $y_j$ on the right
hand side of the correspondence \eqref{OSP} are trivial in imaginary
Liouville theory, i.e.\
$$
\Phi^{(\frac12)}_{- \frac{1}{2b}} (y) \sim V_{-\frac{1}{2b}}(y)\ .
$$
Hence, the imaginary Liouville theory is merely a spectator throughout
most of the computations performed in \cite{Hikida:2008pe}. Consequently,
we can express correlation functions in the OSP(1$|$2) WZW model through
a product of sine-Liouville and imaginary Liouville theory. It then
remains to rewrite the latter in terms of a more conventional theory.
We shall return to these issues in a forthcoming paper. \bigskip

\noindent
{\bf Acknowledgements:} We wish to thank Stefan Fredenhagen, Leszek Hadasz,
Yasuaki Hikida, Zbigniew Jaskolski and J\"org Teschner for useful discussions
and comments. This work was supported in part by the SFB 676.
The work of PS was supported by the Kolumb Programme KOL/6/2011-I of FNP
and by the NCN grant DEC2011/01/B/ST1/01302.

\end{document}